\documentclass[preprint,12pt]{elsarticle}

\usepackage{amssymb}
\usepackage{amsmath}
\usepackage{algorithm}
\usepackage{algorithmic}
\usepackage{array}
\usepackage[caption=false,font=footnotesize,labelfont=sf,textfont=sf]{subfig} 
\usepackage{textcomp}
\usepackage{url}
\usepackage{verbatim}

\usepackage{booktabs} 
\usepackage{tabularx}

\usepackage{tabularx, multirow,multicol,colortbl,array}
\usepackage[dvipsnames]{xcolor}
\usepackage{float}
\usepackage{comment,psfrag}
\usepackage{enumitem}
\usepackage{tikz,tikzscale,bigints,pgfplots}
\pgfplotsset{compat=1.18}
\usetikzlibrary{positioning, arrows.meta, colorbrewer, calc, fit, matrix, external}
\usepgfplotslibrary{fillbetween}

\biboptions{sort&compress}

\journal{Physical Communication}

\begin{document}

\begin{frontmatter}

\title{Turbo Receiver Design for Differentially Encoded PSK in Bursty Impulsive Noise Channels}

\author[ict]{Chin-Hung Chen\corref{cor1}}
\ead{c.h.chen@tue.nl}

\author[cel]{Boris Karanov}
\ead{boris.karanov@kit.edu}

\author[ict,nxp]{Wim van Houtum}
\ead{w.j.v.houtum@tue.nl}

\author[nxp]{Yan Wu}
\ead{yan.wu\_2@nxp.com}

\author[ict]{Alex Alvarado}
\ead{a.alvarado@tue.nl}

\cortext[cor1]{Corresponding author.}

\affiliation[ict]{organization={Information and Communication Theory (ICT) Lab, Dept. Electrical Engineering, Eindhoven University of Technology},
            addressline={},
            city={Eindhoven},
            postcode={5600 MB},
            country={The Netherlands}}

\affiliation[nxp]{organization={NXP Semiconductors},
            addressline={High Tech Campus 60},
            city={Eindhoven},
            postcode={5656 AE},
            country={The Netherlands}}

\affiliation[cel]{organization={Communications Engineering Lab (CEL), Karlsruhe Institute of Technology},
            addressline={},
            city={Karlsruhe},
            postcode={76187},
            country={Germany}}

\begin{abstract}
It has been recognized that the impulsive noise (IN) generated by power devices poses significant challenges to wireless receivers. In this paper, we comprehensively assess the achievable information rate (AIR) for the well-established Markov-Middleton IN model with a phase-shift keying (PSK) input sequence across various channel conditions, including matched and mismatched decoding scenarios. Upon determining information-theoretic bounds, we propose an optimal turbo-differentially encoded (DE)-PSK-IN receiver design based on a commonly used commercial transmission setup consisting of a convolutional encoder, bit-level interleaver, and a DE-PSK symbol mapper. We show that by incorporating the differential decoder into the maximum a-posteriori-based (MAP) IN detector, we can significantly enhance the receiver performance with a $4.5$~dB gain compared to the conventional MAP-based turbo-PSK-IN receiver and a gap of around $1$~dB to the theoretical bounds. We also propose a suboptimal separate receiver design that can be implemented with half the complexity of the joint design and near-optimal performance. We have evaluated the performance of the proposed receiver designs through extensive simulations, demonstrating their effectiveness in real-world scenarios with limited interleaver depth and mismatched state implementation.
\end{abstract}


\begin{highlights}  
\item Unified HMM representation for deriving AIR, MAP detector/demapper/decoder, and iterative decoding
\item Matched and mismatched AIR analysis for Markov–Middleton model to provide practical system design guidance.  
\item Propose an optimal turbo-DE-PSK-IN receiver design that surpasses the performance of a conventional turbo-PSK-IN receiver and approaches near-capacity performance.  
\item Propose a suboptimal turbo receiver design with half the complexity with minor loss. 
\end{highlights}


\end{frontmatter}



\section{Introduction}\label{sec:intro}
Electromagnetic interference (EMI) generated by power devices such as switching gears, power lines, and DC-DC converters presents significant challenges to receivers traditionally designed under the additive white Gaussian noise (AWGN) assumption. Measurements have shown that electronic components in electric vehicles (EVs) \cite{Maouloud21, Gao15,Pliakostathis20,chen2024} and power substations \cite{Shan09, Sacuto14} emit EMI that can be detected by high-sensitivity receivers. The impulsive nature of this interference has been shown to strongly affect powerline communications \cite{Pighi06, Zimmermann02}, digital broadcasting systems \cite{Wim22, Landa15}, and wireless communication receivers within power substations \cite{Shan11,Bhatti10, Sarr17}. 

Conventional clipping \cite{Ndo10} and blanking \cite{Oh17} detectors for IN mitigation are widely adopted due to their simple implementation and the low computational complexity. However, their performance is far from optimal, especially in severe IN channel conditions \cite{Alam20}. To design robust receivers for channels impaired by IN, it is essential to obtain their statistical properties, and therefore, an appropriate IN model is necessary. 
Among the best-known IN models, the Middleton Class A model \cite{Middleton97, Middleton99} provides a tractable probability density function (PDF) based on a Poisson mixture that can be specified by only three parameters. However, it cannot capture the bursty temporal correlations observed in practice. To address this limitation, hidden Markov models (HMMs) have been adopted for IN modeling \cite{Zimmermann02, Dario09, MMA} to address the correlation between impulsive samples. In \cite{Dario09}, a two-state Markov-Gaussian model was proposed, where a comprehensive achievable information rate (AIR) analysis was presented. In \cite{MMA}, the authors combined the HMM process with the Middleton Class A model and proposed the Markov-Middleton model. In \cite{chen2024}, it was shown that a Markov-Middleton model with finite IN states effectively represents the EV-induced interference in practice. However, to the best of our knowledge, an information-theoretical analysis of this model has not been thoroughly investigated; therefore, one of the goals of this paper is to fill this gap by conducting a detailed AIR analysis under various channel conditions for both matched and mismatched decoding scenarios.

Optimal IN detectors exploit the IN statistics to compute maximum a posteriori (MAP) symbol decisions. In \cite{Haring02, Dario07, Dario09_2, MMA, Alam20}, several optimal MAP-based IN detectors have been derived for different IN models. To further enhance the receiver performance, powerful coding systems in combination with the IN detector have been extensively investigated to approach the AIR for the underlying channel model \cite{MMA, Dario09, Mitra10, Mengistu14, Tseng14, Alam20, Nakagawa05, Umehara04}. Receivers for these schemes take the soft information provided by the MAP-based IN detector and pass it to the powerful decoding algorithms, such as low-density parity-check (LDPC) codes \cite{MMA,Dario09, Alam20, Nakagawa05}, convolutional codes \cite{Mitra10, Mengistu14}, and turbo-like codes \cite{Tseng14,Umehara04}. However, previous studies have primarily focused on a basic binary phase-shift keying (BPSK) modulation scheme, mainly addressing the design of its corresponding IN detector and decoder. There has been limited investigation into more complex modulation schemes used in real-world transmission formats for IN receivers.

In wireless communication standards, a differential mapper, where the information is encoded in the phase difference between successive symbols, is commonly used to mitigate the effects of phase ambiguity and for noncoherent detection. For example, digital audio broadcasting, terrestrial digital multimedia broadcasting, and IEEE 802.11g WiFi employ convolutional coding with a differentially encoded (DE)-PSK symbol mapper. Although initially designed for non-coherent detection, research has demonstrated that a coherent DE-PSK demapper can be constructed, wherein the soft information output can be computed using the MAP rule \cite{Colavolpe04}. Since then, turbo-DE-PSK systems, which combine a MAP DE-PSK demapper with a MAP decoder, have been explored in a broad range of literature, see \cite{Narayanan99, Hoeher99, WvH12} and references therein. The recursive nature of the DE-PSK modulator increases the time diversity of a serial turbo-like receiver structure. As a result, a turbo-DE-PSK system achieves significant performance gains compared to a simple turbo-PSK receiver, as reported in \cite{Lee00}.

In this paper, we propose to combine a simple differential demapper with the MAP IN detector, forming a MAP-based joint DE-PSK-IN demapper, which can be implemented with the same complexity as a conventional PSK-IN demapper. We further construct a turbo-DE-PSK-IN receiver structure that significantly outperforms its MAP-based turbo-PSK-IN counterpart. This paper starts with an information-theoretical analysis of the Markov-Middleton channel and then performs turbo receiver designs to approach the estimated theoretical bounds. The main contributions are as follows: (a) We provide a unified HMM representation for deriving AIR, MAP detector/demapper/decoder, and iterative decoding metrics through modified branch metrics. This general representation facilitates a streamlined analysis, reveals the underlying commonalities among different algorithms, and demonstrates that they can be systematically constructed under a similar forward-backward framework. (b) We extend the AIR analysis in \cite{Dario09} to adapt it to a finite-state Markov-Middleton channel with a PSK input sequence. Our comprehensive analysis covers both matched and mismatched decoding scenarios, providing guidance on receiver parameter selections. (c) We propose a novel turbo-DE-PSK-IN receiver design that approaches the theoretical limits of the Markov-Middleton channel within a gap of approximately $1$~ dB, and show that it can achieve a significant performance gain compared to the conventional turbo-PSK-IN receiver proposed in \cite{MMA, Dario09}. The ultimate performance of the proposed receiver with perfect likelihood input is also presented, and (d) we further propose a suboptimal turbo-DE-PSK-IN receiver design that can be implemented with half the complexity of the optimal joint design with near-optimal performance. The proposed receiver with limited interleaver depth and a mismatched state implementation is also investigated and discussed. 
Note that in this study, we assume the IN parameters are perfectly known at the receiver to establish the achievable performance bounds of the proposed turbo-DE-PSK-IN structure. In practice, the accuracy of the estimated IN parameters can significantly influence the MAP detector's reliability. For comprehensive reviews of parameter estimation methods for Middleton Class A and Markov-Gaussian models, readers are referred to  \cite{chc25, Zabin91, Chen19, Vannucci19}.

The paper is organized as follows: Sec.~\ref{sec:hmm} presents the general structure and inference algorithm of HMMs. Sec.~\ref{sec:MMA} reviews the Markov-Middleton IN channel and its HMM representation. Sec.~\ref{sec:inforate} illustrates the computation of the AIR for the considered IN channel with a PSK input sequence. Sec.~\ref{sec:receiver} details the proposed communication system and the design methodology for the optimal joint and suboptimal separate turbo-DPSK-IN receivers. Sec.~\ref{sec:simu} presents the simulation results for AIR and the proposed receiver performance, and finally, Sec.~\ref{sec:conc} concludes this paper. 

\textbf{Notation:} Throughout this paper, we use lowercase letters to represent a scalar (e.g., $s_t$) with subscripts indicating a specific time instant and boldface letters to denote a sequence (e.g., $\mathbf{s}^T_1$), with subscripts and superscripts denoting the start and end of the sequence. The set of all states is denoted by calligraphic letters (e.g., $\mathcal{S}$). For the reader's convenience, the primary symbols and parameters used throughout the paper are summarized in Table~\ref{tab:notation}.

\begin{table}[t!]
    \centering
    \caption{List of main symbols and notations}
    \label{tab:notation}
    \renewcommand{\arraystretch}{1.2}
    \setlength{\tabcolsep}{15pt} 
    \begin{tabularx}{\columnwidth}{@{} l X @{}}
        \toprule
        \textbf{Symbol} & \textbf{Description} \\
        \midrule

       \multicolumn{2}{@{} l}{\textbf{\textit{Transmission Model Parameters}}} \\
        $M$ & Modulation order \\
        $b_k, c_k, d_k$ & Information, coded, and interleaved coded bits \\
        $x_t, z_t$ & PSK-modulated and differentially encoded symbols \\
        $n_t, y_t$ & Noise realization and observation symbols \\
        \addlinespace[0.5em]
        
        \multicolumn{2}{@{} l}{\textbf{\textit{Markov--Middleton Noise Model}}} \\
        $W$ & Number of noise states \\
        $A, \Lambda$ & Impulsive index and impulsive-to-background power ratio \\
        $r$ & Correlation parameter between noise states \\
        $\sigma_0^2, \sigma_j^2$ & Background and state-dependent noise variances \\
        $P'_j$ & Prior state probability of a truncated noise model \\
        $P_{ij}$ & Transition probability of noise states \\
        \addlinespace[0.5em]

        \multicolumn{2}{@{} l}{\textbf{\textit{Hidden State Variables}}} \\
        $s_t$ & Generic hidden state \\
        $w_t$ & Impulsive noise state \\
        $a_t, \dot{a}_t$ & Auxiliary states for PSK-IN and DE-PSK-IN demappers \\
        $l_t$ & Auxiliary state of the differential decoder \\
        $g_k$ & Auxiliary state of the convolutional decoder \\
        \bottomrule
    \end{tabularx}
\end{table}

\section{Hidden Markov Model and its Efficient Inference Algorithm}\label{sec:hmm}
This section provides a general overview of the structure and statistical representation of HMMs. Additionally, we present the utilization of the forward-backward algorithm to compute joint probabilities of observations and latent variables. An HMM structure will be used to describe the Markov-Middleton IN channel in Sec.~\ref{sec:MMA}, and the inference algorithm presented in this section will later be applied to AIR computation in Sec.~\ref{sec:inforate}, while the application to symbol detection, differential demapping, and convolutional decoding based on the MAP rule will be provided in Sec.~\ref{sec:receiver}. 

An HMM is a statistical model that captures the temporal behavior of a sequence of latent variables (states) $\mathbf{s}^T_1 = (s_1, s_2, \cdots, s_{T})$ from a Markov process via a sequence of state-dependent observable variables (observations) $\mathbf{y}^T_1 = (y_1, y_2, \cdots, y_{T})$. Here, $s_t$ denotes the state realization at time $t$, which is drawn from a time-invariant finite-state set 
\begin{align}
    s_t \in \mathcal{S} \triangleq \{0,1,\cdots,|\mathcal{S}|-1\},    
\end{align}
where $|\mathcal{S}|$ is the size of the state space. A general representation of the structure of an HMM is shown in Fig.~\ref{fig:hmm} with the transition probability $p(s_t = j \mid s_{t-1}=i)$ representing the probability of moving from state $i$ at time $t-1$ to state $j$ at time $t$. The observation likelihood $p(y_t\mid s_t=j)$, on the other hand, describes the probability of an observation being generated from a state $j$ at time $t$. 

\begin{figure}[t]
    \centering
    {\begin{tikzpicture}[
    node distance = 1.0cm and 1.6cm,
    state/.style={shape=circle, draw=Black, fill=Black!40, minimum size=1cm, thick,font=\small},
    obs/.style={shape=circle, draw=Black,, fill=ForestGreen!40, minimum size=1cm, thick,font=\small},
    every edge/.style={draw, -stealth, thick},
    label/.style={above, text height=1.5ex, text depth=.25ex},
    obslabel/.style={right, text height=1.5ex, text depth=.25ex},
]
\def\ttsize{\small}
\node[state] (s_t-1) at (0, 0) {$s_{t-1}$};
\node[state] (s_t)   [right=of s_t-1] {$s_t$};
\node[state] (s_t+1) [right=of s_t] {$s_{t+1}$};

\node[obs] (y_t-1) at ($(s_t-1) - (0,2)$) {$y_{t-1}$};
\node[obs] (y_t)   [below=of s_t] {$y_t$};
\node[obs] (y_t+1) [below=of s_t+1] {$y_{t+1}$};

\path (s_t-1) edge node[label] {\ttsize$p(s_t|s_{t-1})$} (s_t);
\path (s_t) edge node[label] {\ttsize$p(s_{t+1}|s_t)$} (s_t+1);
\path (s_t-1) edge node[obslabel] {\ttsize$p(y_{t-1}|s_{t-1})$} (y_t-1);
\path (s_t) edge node[obslabel] {\ttsize$p(y_t|s_t)$} (y_t);
\path (s_t+1) edge node[obslabel] {\ttsize$p(y_{t+1}|s_{t+1})$} (y_t+1);

\node at ($(s_t-1) - (1,0)$) {\dots};
\node at ($(s_t+1) + (1,0)$) {\dots};

\end{tikzpicture}} 
    \caption{Structure of a hidden Markov model, showing the factorization and conditional independencies implied by the model.}   
    \label{fig:hmm}
\end{figure}

The Markovian structure of an HMM allows an efficient inference of the hidden states through the well-known Bahl-Cocke-Jelinek-Raviv (BCJR) algorithm \cite{bcjr74}. Specifically, the inference problem aims at calculating the posterior belief of a latent state $s_t$ at a specific time instant given the full sequence of observations as
\begin{equation} \label{eq:map}
    p(s_t \mid \mathbf{y}^T_1) \propto p(s_t, \mathbf{y}^T_1) = \sum_{s_{t-1}\in\mathcal{S}}{p(s_{t}, s_{t-1}, \mathbf{y}^T_1)}.
\end{equation}
The right-most side of \eqref{eq:map} represents the joint distribution of the state pairs $(s_t,s_{t-1})$ at a specific time $t$ and the whole received sequence $\mathbf{y}_1^T$. We can further factorize the joint probability as
\begin{align} \label{eq:bcjr}
    p(s_t, s_{t-1},\mathbf{y}^T_1) = \underbrace{p(\mathbf{y}_{t+1}^{T} \mid s_{t})}_{\beta(s_t)} \cdot \underbrace{p(y_t, s_t \mid s_{t-1})}_{\gamma(s_t,s_{t-1})}  \cdot \underbrace{p(s_{t-1}, \mathbf{y}_1^{t-1})}_{\alpha(s_{t-1})}. 
\end{align}

In \eqref{eq:bcjr}, $\alpha(s_{t-1})$, $\beta(s_t)$, and $\gamma(s_t, s_{t-1})$ represent $p(s_{t-1}, \mathbf{y}_1^{t-1})$, $p(\mathbf{y}_{t+1}^{T} \mid s_{t})$, and $p(y_t, s_t \mid s_{t-1})$, respectively, corresponding to forward recursion, backward recursion, and branch metrics.

Using Bayes' rule and the Markovian property, the branch metric $\gamma$ can be further derived as 
\begin{align} \label{eq:gamma}
        \gamma(s_{t}, s_{t-1}) \triangleq p(y_t, s_t \mid s_{t-1}) =  p(y_t \mid s_t) \cdot p(s_{t} \mid s_{t-1})
\end{align}
which can be seen as the multiplication of the observation likelihood and the state transition probability.
The forward recursion representation of $\alpha$ and the backward recursion $\beta$ can be written as
\begin{align} \label{eq:siso_alpha}
    \alpha(s_{t}) &\triangleq p(s_{t}, \mathbf{y}_1^{t}) = \sum_{s_{t-1}\in\mathcal{S}} \gamma(s_{t}, s_{t-1}) \cdot \alpha(s_{t-1}),
\end{align}
and
\begin{align} \label{eq:siso_beta}
    \beta(s_{t-1}) &\triangleq p(\mathbf{y}_{t}^{T} \mid s_{t-1}) = \sum_{s_{t}\in\mathcal{S}} \beta(s_t) \cdot \gamma(s_{t}, s_{t-1}).
\end{align}

The forward-backward computation \eqref{eq:map}--\eqref{eq:siso_beta} of the posterior belief comprises the conventional BCJR algorithm (see \cite{bcjr74} for detailed derivations). In this paper, we will examine several adaptations of this algorithm, each utilizing the same forward-backward framework but with modifications to the branch metrics $\gamma$ that are tailored to specific applications.

\section{Markov--Middleton Impulsive Noise Model}\label{sec:MMA}
\begin{figure}[t]
\centering
    \subfloat[$A=0.1, \Lambda = 10, r=0$\label{fig:MMA_noise1}]{
        \resizebox{0.45\linewidth}{!}{\begin{tikzpicture}
    \begin{axis}[ 
        axis y line*=right,
        axis line style = thick,
        axis x line=none,
		grid = both,
		name = p1,
		width=\columnwidth,
		height=.3\columnwidth, 
  		xmin = 200, xmax=500,
	   	ymin = .8, ymax=4,
  		ytick = {1,2,3,4}, yticklabels={0,1,2,3},
            ylabel = State,
            ylabel style={color=orange},
            yticklabel style={color=orange}, 
		font=\small,
        ]
		\addplot[ycomb, very thin, color=orange!50, mark=*, mark size=1, mark options={fill=white, semithick}] 
		table[x expr=\thisrowno{0}, y expr=\thisrowno{2}] 
		{noise_A01G10r0.txt};
	\end{axis}
 
	\begin{axis}[ 
        axis line style = thick,
		name = p1,
		width=\columnwidth,
		height=.3\columnwidth, 
  		xmin = 200, xmax=500,
	   	ymin = 0, ymax=25,
            tick style={draw=none},
  		ytick = {0,10,...,20}, yticklabels={0,5,...,20},
            xtick = {200,300,...,1000}, xticklabels={0,100,...,1000},
            xlabel = Sample, ylabel = Mag.,
		font=\small,
 ]
		\addplot[color=black, semithick] 
		table[x expr=\thisrowno{0}, y expr=\thisrowno{1}] 
		{noise_A01G10r0.txt};
		
	\end{axis}
\end{tikzpicture}    }
    }
    \hfill
    \subfloat[$A=0.3, \Lambda = 10, r=0$\label{fig:MMA_noise2}]{
        \resizebox{0.45\linewidth}{!}{	
\begin{tikzpicture}
    \begin{axis}[ 
        axis y line*=right,
        axis line style = thick,
        axis x line=none,
		grid = both,
		name = p1,
		width=\columnwidth,
		height=.3\columnwidth, 
  		xmin = 200, xmax=500,
	   	ymin = .8, ymax=4,
  		ytick = {1,2,3,4}, yticklabels={0,1,2,3},
            ylabel = State,
            ylabel style={color=orange},
            yticklabel style={color=orange}, 
		font=\small,
        ]
        \addplot[ycomb, very thin, color=orange!50, mark=*, mark size=1,mark options={fill=white, semithick}] 
		table[x expr=\thisrowno{0}, y expr=\thisrowno{2}] 
		{noise_A03G10r0.txt};
        \addplot[name path=B, fill=none] {0};

	\end{axis}
 
	\begin{axis}[ 
        axis line style = thick,
		name = p1,
		width=\columnwidth,
		height=.3\columnwidth, 
  		xmin = 200, xmax=500,
	   	ymin = 0, ymax=25,
            tick style={draw=none},
  		ytick = {0,10,...,20}, yticklabels={0,5,...,20},
            xtick = {200,300,...,1000}, xticklabels={0,100,...,1000},
            xlabel = Sample, ylabel = Mag.,
		font=\small,
 ]
		\addplot[color=black, semithick] 
		table[x expr=\thisrowno{0}, y expr=\thisrowno{1}] 
		{noise_A03G10r0.txt};
		
	\end{axis}

\end{tikzpicture}    }
    }\\[1ex]
    \subfloat[$A=0.3, \Lambda = 50, r=0$\label{fig:MMA_noise3}]{
        \resizebox{0.45\linewidth}{!}{	
\begin{tikzpicture}
	\begin{axis}[ 
        axis y line*=right,
        axis line style = thick,
        axis x line=none,
		grid = both,
		name = p1,
		width=\columnwidth,
		height=.3\columnwidth, 
  		xmin = 200, xmax=500,
	   	ymin = .8, ymax=4,
  		ytick = {1,2,3,4}, yticklabels={0,1,2,3},
            ylabel = State,
            ylabel style={color=orange},
            yticklabel style={color=orange}, 
		font=\small,
        ]
		\addplot[ycomb, very thin, color=orange!50, mark=*, mark size=1,mark options={fill=white, semithick}] 
		table[x expr=\thisrowno{0}, y expr=\thisrowno{2}] 
		{noise_A03G100r0.txt};
    \end{axis}
  
	\begin{axis}[ 
        axis line style = thick,
		name = p1,
		width=\columnwidth,
		height=.3\columnwidth, 
  		xmin = 200, xmax=500,
	   	ymin = 0, ymax=25,
            tick style={draw=none},
  		ytick = {0,10,...,20}, yticklabels={0,5,...,20},
            xtick = {200,300,...,1000}, xticklabels={0,100,...,1000},
            xlabel = Sample, ylabel = Mag.,
		font=\small,
        ]
		\addplot[color=black, semithick] 
		table[x expr=\thisrowno{0}, y expr=\thisrowno{1}] 
		{noise_A03G100r0.txt};
		
	\end{axis}

\end{tikzpicture}    }
    }
    \hfill
    \subfloat[$A=0.3, \Lambda = 50, r=0.9$\label{fig:MMA_noise4}]{
        \resizebox{0.45\linewidth}{!}{	
\begin{tikzpicture}
 
	\begin{axis}[ 
        axis y line*=right,
        axis line style = thick,
        axis x line=none,
		grid = both,
		name = p1,
		width=\columnwidth,
		height=.3\columnwidth, 
  		xmin = 200, xmax=500,
	   	ymin = .8, ymax=4,
  		ytick = {1,2,3,4}, yticklabels={0,1,2,3},
            ylabel = State,
            ylabel style={color=orange},
            yticklabel style={color=orange}, 
		font=\small,
        ]
		\addplot[ycomb, very thin, color=orange!50, mark=*, mark size=1, mark options={fill=white, semithick}] 
		table[x expr=\thisrowno{0}, y expr=\thisrowno{2}] 
		{noise_A03G100r9.txt};
		
	\end{axis}

	\begin{axis}[ 
            axis line style = thick,
		name = p1,
		width=\columnwidth,
		height=.3\columnwidth, 
  		xmin = 200, xmax=500,
	   	ymin = 0, ymax=25,
            tick style={draw=none},
  		ytick = {0,10,...,20}, yticklabels={0,5,...,20},
            xtick = {200,300,...,1000}, xticklabels={0,100,...,1000},
            xlabel = Sample, ylabel = Mag.,
		font=\small,
        ]
		\addplot[color=black, semithick] 
		table[x expr=\thisrowno{0}, y expr=\thisrowno{1}] 
		{noise_A03G100r9.txt};
		
	\end{axis}
 
\end{tikzpicture}    }
    }
    \caption{Noise realization from a $4$-state Markov--Middleton impulsive noise model with a background noise variance $\sigma_0^2=1$ for impulsive index $A=0.1$ and $A=0.3$, impulsive-to-background noise power ratio $\Lambda=10$ and $\Lambda=50$, and correlation parameters $r=0$ and $ r=0.9$. The orange vertical bar denotes the index $j$ of the state realizations \( w_t = j \) for generating the corresponding noise samples.}
    \label{fig:MMA_noise}
\end{figure}
In this section, we summarize the Markov-Middleton model, initially developed for powerline communications~\cite{MMA}, which captures the statistical behavior of bursty IN by combining the conventional Middleton Class A model~\cite{Middleton97} with an HMM. The integration of HMM introduces memory between noise samples, with state transition probabilities characterizing the correlation between consecutive bursts.

The Markov-Middleton model is characterized by four parameters: the background noise (BN) variance $\sigma_0^2$, the impulsive index $A$, which quantifies the probability of impulsive (non-BN) events, the impulsive-to-background noise power ratio $\Lambda$, which determines the intensity of IN compared to BN, and the noise correlation parameter $r$, which describes the correlation between consecutive noise samples. The noise realizations generated by the Markov-Middleton model for different parameters are depicted in Figs.~\ref{fig:MMA_noise1}--\ref{fig:MMA_noise4}, and will be further explained in the following.

The statistical properties of the IN samples $n_t$ are completely determined by the noise state realization at time $t$, which is defined as 
\begin{align}\label{eq:state_mma}
    w_t \in \mathcal{W} \triangleq \{0,1,\cdots, |\mathcal{W}|-1\}, 
\end{align}
where $w_t=0$ represents the BN state and $|\mathcal{W}|=W$ is the total number of noise states. 
The PDF of $n_t$ conditioned on a specific noise state $w_t=j$ follows a complex Gaussian distribution $\mathcal{CN}(0,\sigma_j^2)$, which can be expressed as
\begin{align}\label{eq:state_noise}
    p(n_t\mid w_t = j) = \frac{1}{{\pi \sigma^2_j}}\exp{\left(-\frac{|n_t|^2}{\sigma^2_j}\right)},
\end{align}
with $\sigma_j^2$ denote the variance of the complex value noise sample $n_t$ being at a specific state $w_t=j$.

In the original Middleton Class A model \cite{Middleton97}, the number of noise states is assumed to be infinity ($W\rightarrow \infty$). Therefore, the generic PDF of the Middleton Class A noise sample can be written as the expectation over all possible states as
\begin{align}\label{eq:MMApdf}
    p(n_t) &= \sum_{j=0}^{\infty} P_j \cdot p(n_t\mid w_t = j) \nonumber \\ 
           &= \sum_{j=0}^{\infty} \frac{P_j}{{\pi \sigma^2_j}}\exp{\left(-\frac{|n_t|^2}{\sigma^2_j}\right)},
\end{align}
where 
\begin{align}\label{eq:pm}
    P_j = p(w_t = j) = \exp(-A)A^j/j!
\end{align}
represents the prior probability of being at a specific noise state. From \eqref{eq:pm}, we can observe that when $j=0$, the probability of being at BN state $w_t=0$ is $P_B = P_0 = \exp(-A)$ and the probability of being at any of the impulsive states $j>0$ is $P_I = 1 - \exp(-A)$. In other words, a larger impulsive index $A$ results in a larger $P_I$, and thus, a larger proportion of samples with large variance ($\sigma^2_j,\quad j>0$) are generated in a noise sequence $\mathbf{n}_1^T$. As shown in Figs.~\ref{fig:MMA_noise1} and \ref{fig:MMA_noise2}, when the impulsive index $A$ is increased from $0.1$ to $0.3$, the number of noise in IN states ($j>0$) increases.

As illustrated in \eqref{eq:pm}, the probability of being in the $j$-th noise state decreases as $j$ increases. Therefore, it is sufficient to use a truncated Middleton Class A model with a finite number of states $W$ to represent the IN channel with
\begin{align}\label{eq:MMApdf2}
    p'(n_t) = \sum_{j=0}^{W-1} \frac{P'_j}{{\pi \sigma^2_j}}\exp{\left(-\frac{|n_t|^2}{\sigma^2_j}\right)},
\end{align}
where the probability of being in state $j$ is now written as
\begin{align}\label{eq:pm2}
    P'_j = \frac{P_j}{\sum_{i=0}^{W-1}P_i}.
\end{align}
The denominator in \eqref{eq:pm2} is a normalization factor to ensure the total probability equals $1$.

The noise variance in \eqref{eq:MMApdf2} is defined as
\begin{align}\label{eq:varm}
    \sigma^2_j = \left( 1 + \frac{j\Lambda}{A} \right) \sigma_0^2,
\end{align}
where 
\begin{align}\label{eq:lambda}
    \Lambda = \sigma_I^2/\sigma_B^2
\end{align}
is the impulsive-to-background average noise power ratio.\footnote{In the original paper \cite{Middleton97}, the power ratio is defined as $\Gamma=\sigma_B^2/\sigma_I^2$.} The average BN power is $\sigma_B^2 = P'_0 \sigma_0^2$ and the average IN power $\sigma_I^2$ can be written as
\begin{align}
    \sigma_I^2 = \sum_{j=1}^{W-1}{P'_j \sigma_j^2}. \nonumber
\end{align}
From \eqref{eq:lambda}, we can see that as the impulsive-to-background noise ratio $\Lambda$ increases, the difference between the IN power and the BN power increases. This is clearly shown by comparing the noise realizations (black lines) in Figs.~\ref{fig:MMA_noise2} and \ref{fig:MMA_noise3}.

To describe the burstiness (memory) of the noise observed in practice, the Markov-Middleton model \cite{MMA} introduces a parameter $r\in[0, 1]$ to establish a correlation between consecutive noise samples. Specifically, the state transition matrix of the Markov-Middleton noise model is constructed as 
\begin{align}\label{eq:P}
    P_{ij} &\triangleq p(w_{t}=j \mid  w_{t-1} = i) \nonumber \\
    &= \begin{cases}
        r + (1-r) \cdot P'_j , & i=j\\
        (1-r) \cdot P'_j, & \text{otherwise}.
    \end{cases}
\end{align}
From \eqref{eq:P}, we can see that the larger the correlation parameter $r$, the more likely the following noise sample will stay in the same state. Namely, the bursty behavior is more prominent. In Fig.~\ref{fig:MMA_noise4}, the noise samples exhibit strong correlations ($r=0.9$), where the noise tends to remain in the same state (vertical orange bars) for consecutive samples. In contrast, in Figs.~\ref{fig:MMA_noise1}--\ref{fig:MMA_noise3}, the noise samples are randomly chosen from $w_t \in \mathcal{W}$ with probability $P'_j$ defined in \eqref{eq:pm2}. Note that when $r=0$, the Markov-Middleton model is reduced to a truncated memoryless Middleton Class A channel.

\section{Achievable Information Rate}\label{sec:inforate}
This section presents an information-theoretical analysis for transmission over channels with IN following the Markov-Middleton model introduced in Sec.~\ref{sec:MMA}. This is an essential reference for practical receiver design operating in specific channel conditions. We use a simulation-based method to calculate the AIR between the PSK-modulated input sequence $\mathbf{x}_1^T$ and the channel output sequence $\mathbf{y}_1^T$ with an input-output relationship defined as
\begin{align}\label{eq:airio}
    y_t = x_t + n_t,
\end{align}
where 
\begin{align}\label{eq:xsym}
    x_t\in \mathcal{X}=\{e^{j2\pi i/M} \mid i=0,1,\cdots,M-1 \}
\end{align}
denotes the transmitted $M$-ary PSK symbol and $n_t$ follows the Markov-Middleton noise model described in Sec.~\ref{sec:MMA}, which is drawn from the distribution defined in \eqref{eq:MMApdf2}. The computation of AIR for channels with memory is explained in detail in \cite{Arnold06}, and its application to a two-state Markov-Gaussian channel is described in \cite{Dario09}. The algorithm and the computation steps are briefly summarized as follows.

An AIR is the number of information bits that can be reliably transmitted and decoded, with an arbitrarily low error probability per channel use, for a specific modulation format. Specifically, the AIR we consider in this work is the mutual information $I(\mathbf{X};\mathbf{Y})$ between the discrete input random process $\mathbf{X}$ and the continuous output random process $\mathbf{Y}$, which can be estimated via
\begin{align}\label{eq:Iest}
    I(\mathbf{X};\mathbf{Y}) \approx \hat{I}(\mathbf{X};\mathbf{Y}) = \frac{1}{T}\log{p(\mathbf{y}_1^T \mid  \mathbf{x}_1^T)} - \frac{1}{T}\log{p(\mathbf{y}_1^T)} 
\end{align}
with a very long sequence $T$. To calculate \eqref{eq:Iest} for channels with memory, \cite{Arnold06} introduces a simulation-based computation through a sum-product algorithm with the help of an auxiliary state process. In this paper, we define the auxiliary state realization for a PSK-modulated symbol over a Markov-Middleton IN channel as
\begin{align}\label{eq:state_qpskin}
     a_t \triangleq f(x_t\in \mathcal{X},w_t\in \mathcal{W}) \in \mathcal{A} \triangleq \{0,1,\cdots,|\mathcal{A}|-1\},
\end{align}
where $f:\mathcal{X}\times\mathcal{W}\rightarrow \mathcal{A}$ maps each pair $(x_t,w_t)$ in the product space $\mathcal{X} \times \mathcal{W}$ into a unique state in $\mathcal{A}$. The size of the auxiliary state space is thus $|\mathcal{A}| =|\mathcal{X}|\times|\mathcal{W}|=M\times W$. In Fig.~\ref{fig:trellis_IN2QPSK}, we show an example of the trellis section and the auxiliary state for a QPSK transmission (i.e., $M=4$) over a $2$-state (i.e., $W=2$) Markov-Middleton channel. 

The transition probability of the auxiliary states considers the prior PSK symbol probability $p(x_t)$ and the noise state transition matrix defined in \eqref{eq:P}, which can be expressed as
\begin{align} \label{eq:P_dotstran}
    p(a_t \mid  &a_{t-1}) = P_{ij} \cdot p(x_t), \quad w_t=j, w_{t-1} = i.
\end{align}
The channel likelihood function concerning the auxiliary states \eqref{eq:state_qpskin} can be written as
\begin{align}\label{eq:P_dotslik}
	p(y_t\mid a_t) =\frac{1} {{\pi \sigma^2_{j}}}\exp\left(-\frac{|y_{t}-x_t|^2}{\sigma^2_{j}}\right), \quad w_t = j.
\end{align}

\begin{figure}[t] 
    \centering 
    \subfloat[QPSK-IN demapper\label{fig:trellis_IN2QPSK}]
    { \resizebox{0.5\linewidth}{!}{\tikzstyle{state}=[shape=circle,draw=black!50,fill=gray!50, font=\small]
\tikzstyle{observation}=[shape=rectangle,draw=orange!50,fill=orange!20]

\tikzstyle{edge1}=[stealth-,thick, Black]
\tikzstyle{edge2}=[stealth-,thick, Cyan]
\tikzstyle{edge3}=[stealth-,thick, Orange]
\tikzstyle{edge4}=[stealth-,thick, Green]
\tikzstyle{edge5}=[stealth-,thick, Black, dotted]
\tikzstyle{edge6}=[stealth-,thick, Cyan, dotted]
\tikzstyle{edge7}=[stealth-,thick, Orange, dotted]
\tikzstyle{edge8}=[stealth-,thick, Green, dotted]

\tikzstyle{mainstate}=[state,very thick]

\begin{tikzpicture}[]

\def\dist{3}  
\def\v_dist{.8}  

\node  (s') at (0, 5.8*\v_dist)  {$a_{t-1}$};
\node at (-1,5*\v_dist) {\scriptsize$f(e^{j0},0)$};
\node at (-1,4*\v_dist) {\scriptsize$f(e^{j0},1)$};
\node at (-1,3*\v_dist) {\scriptsize$f(e^{j\frac{\pi}{2}},0)$};
\node at (-1,2*\v_dist) {\scriptsize$f(e^{j\frac{\pi}{2}},1)$};
\node at (-1,1*\v_dist) {\scriptsize$f(e^{j\pi},0)$};
\node at (-1,0*\v_dist) {\scriptsize$f(e^{j\pi},1)$};
\node at (-1,-1*\v_dist) {\scriptsize$f(e^{j\frac{3\pi}{2}},0)$};
\node at (-1,-2*\v_dist) {\scriptsize$f(e^{j\frac{3\pi}{2}},1)$};

\node[mainstate] (s1_1) at (0, 5*\v_dist) {$0$};
\node[mainstate] (s2_1) at (0, 4*\v_dist) {$1$};
\node[mainstate] (s3_1) at (0, 3*\v_dist) {$2$};
\node[mainstate] (s4_1) at (0, 2*\v_dist) {$3$};
\node[mainstate] (s5_1) at (0, 1*\v_dist) {$4$};
\node[mainstate] (s6_1) at (0, 0*\v_dist) {$5$};
\node[mainstate] (s7_1) at (0,-1*\v_dist) {$6$};
\node[mainstate] (s8_1) at (0,-2*\v_dist) {$7$};

\node [right=\dist of s'] {$a_{t}$};
\node[mainstate] (s1_2) [right=\dist of s1_1] {$0$}
    edge[edge1] (s1_1)
    edge[edge1] (s3_1)
    edge[edge1] (s5_1)
    edge[edge1] (s7_1);
\node[mainstate] (s2_2) [right=\dist of s2_1] {$1$}
    edge[edge5] (s1_1)
    edge[edge5] (s3_1)
    edge[edge5] (s5_1)
    edge[edge5] (s7_1);
\node[mainstate] (s3_2) [right=\dist of s3_1] {$2$}
    edge[edge2] (s1_1)
    edge[edge2] (s3_1)
    edge[edge2] (s5_1)
    edge[edge2] (s7_1);
\node[mainstate] (s4_2) [right=\dist of s4_1] {$3$}
    edge[edge6] (s1_1)
    edge[edge6] (s3_1)
    edge[edge6] (s5_1)
    edge[edge6] (s7_1);
\node[mainstate] (s5_2) [right=\dist of s5_1] {$4$}
    edge[edge3] (s1_1)
    edge[edge3] (s3_1)
    edge[edge3] (s5_1)
    edge[edge3] (s7_1);
\node[mainstate] (s6_2) [right=\dist of s6_1] {$5$}
    edge[edge7] (s1_1)
    edge[edge7] (s3_1)
    edge[edge7] (s5_1)
    edge[edge7] (s7_1);
\node[mainstate] (s7_2) [right=\dist of s7_1] {$6$}
    edge[edge4] (s1_1)
    edge[edge4] (s3_1)
    edge[edge4] (s5_1)
    edge[edge4] (s7_1);
\node[mainstate] (s8_2) [right=\dist of s8_1] {$7$}
    edge[edge8] (s1_1)
    edge[edge8] (s3_1)
    edge[edge8] (s5_1)
    edge[edge8] (s7_1);

\node[
    anchor=west,
    fill=white,
    font=\footnotesize
    ] at (s4_2.east) {
    \begin{tabular}{@{} c@{}c@{}}
        \multicolumn{2}{c}{Trellis Input} \\ 
        \begin{tikzpicture}
            \draw[edge1,-stealth] (.4,0) -- (.8,0);
        \end{tikzpicture} & \raisebox{-0.3ex}{$(e^{j0}, 0)$} \\
        \begin{tikzpicture}
            \draw[edge5,-stealth] (.4,0) -- (.8,0);
        \end{tikzpicture} & \raisebox{-0.3ex}{$(e^{j0}, 1)$} \\
                \begin{tikzpicture}
            \draw[edge2,-stealth] (.4,0) -- (.8,0);
        \end{tikzpicture} & \raisebox{-0.3ex}{$(e^{j\frac{\pi}{2}}, 0)$} \\
                \begin{tikzpicture}
            \draw[edge6,-stealth] (.4,0) -- (.8,0);
        \end{tikzpicture} & \raisebox{-0.3ex}{$(e^{j\frac{\pi}{2}}, 1)$} \\
                \begin{tikzpicture}
            \draw[edge3,-stealth] (.4,0) -- (.8,0);
        \end{tikzpicture} & \raisebox{-0.3ex}{$(e^{j\pi}, 0)$} \\
                \begin{tikzpicture}
            \draw[edge7,-stealth] (.4,0) -- (.8,0);
        \end{tikzpicture} & \raisebox{-0.3ex}{$(e^{j\pi}, 1)$} \\
                \begin{tikzpicture}
            \draw[edge4,-stealth] (.4,0) -- (.8,0);
        \end{tikzpicture} & \raisebox{-0.3ex}{$(e^{j\frac{3\pi}{2}}, 0)$} \\
                \begin{tikzpicture}
            \draw[edge8,-stealth] (.4,0) -- (.8,0);
        \end{tikzpicture} & \raisebox{-0.3ex}{$(e^{j\frac{3\pi}{2}}, 1)$} \\
    \end{tabular}
};

\end{tikzpicture}}}
    \subfloat[DE-QPSK-IN demapper\label{fig:trellis_IN2DQPSK}]
    { \resizebox{0.5\linewidth}{!}{\tikzstyle{state}=[shape=circle,draw=black!50,fill=gray!50, font=\small]
\tikzstyle{observation}=[shape=rectangle,draw=orange!50,fill=orange!20]

\tikzstyle{edge1}=[stealth-,thick, Black]
\tikzstyle{edge2}=[stealth-,thick, Cyan]
\tikzstyle{edge3}=[stealth-,thick, Orange]
\tikzstyle{edge4}=[stealth-,thick, Green]
\tikzstyle{edge5}=[stealth-,thick, Black, dotted]
\tikzstyle{edge6}=[stealth-,thick, Cyan, dotted]
\tikzstyle{edge7}=[stealth-,thick, Orange, dotted]
\tikzstyle{edge8}=[stealth-,thick, Green, dotted]

\tikzstyle{mainstate}=[state,very thick]

\begin{tikzpicture}[]

\def\dist{3}  
\def\v_dist{.8}  

\node  (s') at (0, 5.8*\v_dist)  {$\dot{a}_{t-1}$};
\node at (-1,5*\v_dist) {\scriptsize$f(e^{j0},0)$};
\node at (-1,4*\v_dist) {\scriptsize$f(e^{j0},1)$};
\node at (-1,3*\v_dist) {\scriptsize$f(e^{j\frac{\pi}{2}},0)$};
\node at (-1,2*\v_dist) {\scriptsize$f(e^{j\frac{\pi}{2}},1)$};
\node at (-1,1*\v_dist) {\scriptsize$f(e^{j\pi},0)$};
\node at (-1,0*\v_dist) {\scriptsize$f(e^{j\pi},1)$};
\node at (-1,-1*\v_dist) {\scriptsize$f(e^{j\frac{3\pi}{2}},0)$};
\node at (-1,-2*\v_dist) {\scriptsize$f(e^{j\frac{3\pi}{2}},1)$};

\node[mainstate] (s1_1) at (0, 5*\v_dist) {$0$};
\node[mainstate] (s2_1) at (0, 4*\v_dist) {$1$};
\node[mainstate] (s3_1) at (0, 3*\v_dist) {$2$};
\node[mainstate] (s4_1) at (0, 2*\v_dist) {$3$};
\node[mainstate] (s5_1) at (0, 1*\v_dist) {$4$};
\node[mainstate] (s6_1) at (0, 0*\v_dist) {$5$};
\node[mainstate] (s7_1) at (0,-1*\v_dist) {$6$};
\node[mainstate] (s8_1) at (0,-2*\v_dist) {$7$};

\node [right=\dist of s'] {$\dot{a}_{t}$};
\node[mainstate] (s1_2) [right=\dist of s1_1] {$0$}
    edge[edge1] (s1_1)
    edge[edge4] (s3_1)
    edge[edge3] (s5_1)
    edge[edge2] (s7_1);
\node[mainstate] (s2_2) [right=\dist of s2_1] {$1$}
    edge[edge5] (s1_1)
    edge[edge8] (s3_1)
    edge[edge7] (s5_1)
    edge[edge6] (s7_1);
\node[mainstate] (s3_2) [right=\dist of s3_1] {$2$}
    edge[edge2] (s1_1)
    edge[edge1] (s3_1)
    edge[edge4] (s5_1)
    edge[edge3] (s7_1);
\node[mainstate] (s4_2) [right=\dist of s4_1] {$3$}
    edge[edge6] (s1_1)
    edge[edge5] (s3_1)
    edge[edge8] (s5_1)
    edge[edge7] (s7_1);
\node[mainstate] (s5_2) [right=\dist of s5_1] {$4$}
    edge[edge3] (s1_1)
    edge[edge2] (s3_1)
    edge[edge1] (s5_1)
    edge[edge4] (s7_1);
\node[mainstate] (s6_2) [right=\dist of s6_1] {$5$}
    edge[edge7] (s1_1)
    edge[edge6] (s3_1)
    edge[edge5] (s5_1)
    edge[edge8] (s7_1);
\node[mainstate] (s7_2) [right=\dist of s7_1] {$6$}
    edge[edge4] (s1_1)
    edge[edge3] (s3_1)
    edge[edge2] (s5_1)
    edge[edge1] (s7_1);
\node[mainstate] (s8_2) [right=\dist of s8_1] {$7$}
    edge[edge8] (s1_1)
    edge[edge7] (s3_1)
    edge[edge6] (s5_1)
    edge[edge5] (s7_1);

\node[
    anchor=west,
    fill=white,
    font=\footnotesize
    ] at (s4_2.east) {
    \begin{tabular}{@{} c@{}c@{}}
        \multicolumn{2}{c}{Trellis Input} \\ 
        \begin{tikzpicture}
            \draw[edge1,-stealth] (.4,0) -- (.8,0);
        \end{tikzpicture} & \raisebox{-0.3ex}{$(e^{j0}, 0)$} \\
        \begin{tikzpicture}
            \draw[edge5,-stealth] (.4,0) -- (.8,0);
        \end{tikzpicture} & \raisebox{-0.3ex}{$(e^{j0}, 1)$} \\
                \begin{tikzpicture}
            \draw[edge2,-stealth] (.4,0) -- (.8,0);
        \end{tikzpicture} & \raisebox{-0.3ex}{$(e^{j\frac{\pi}{2}}, 0)$} \\
                \begin{tikzpicture}
            \draw[edge6,-stealth] (.4,0) -- (.8,0);
        \end{tikzpicture} & \raisebox{-0.3ex}{$(e^{j\frac{\pi}{2}}, 1)$} \\
                \begin{tikzpicture}
            \draw[edge3,-stealth] (.4,0) -- (.8,0);
        \end{tikzpicture} & \raisebox{-0.3ex}{$(e^{j\pi}, 0)$} \\
                \begin{tikzpicture}
            \draw[edge7,-stealth] (.4,0) -- (.8,0);
        \end{tikzpicture} & \raisebox{-0.3ex}{$(e^{j\pi}, 1)$} \\
                \begin{tikzpicture}
            \draw[edge4,-stealth] (.4,0) -- (.8,0);
        \end{tikzpicture} & \raisebox{-0.3ex}{$(e^{j\frac{3\pi}{2}}, 0)$} \\
                \begin{tikzpicture}
            \draw[edge8,-stealth] (.4,0) -- (.8,0);
        \end{tikzpicture} & \raisebox{-0.3ex}{$(e^{j\frac{3\pi}{2}}, 1)$} \\
    \end{tabular}
};

\end{tikzpicture}} } 
    \caption{Trellis sections of an impulsive noise detector with a 2-state Markov--Middleton model for (a) QPSK and (b) DE-QPSK transmissions, respectively. For clarity, we only show the state transitions from states 0, 2, 4, and 6.} 
    \label{fig:trellis_IN2} 
\end{figure}

After determining the auxiliary states with their statistical representation \eqref{eq:P_dotstran} and \eqref{eq:P_dotslik}, the PDF of the channel output sequence $p(\mathbf{y}_1^T)$ can be calculated via
\begin{align}\label{eq:I_py}
    p(\mathbf{y}_1^T) = \sum_{{a}_T \in \mathcal{A}}{p(a_T,\mathbf{y}_1^T)} =  \sum_{{a}_T \in \mathcal{A}}{\alpha(a_T)},
\end{align}
using forward recursion \eqref{eq:siso_alpha}, with a branch metric
\begin{align}\label{eq:gamma_sdot}
    \gamma_a(a_t,a_{t-1}) \triangleq p(y_t \mid  a_t) \cdot  p(a_t \mid  a_{t-1}). 
\end{align}
To compute the PDF of the channel output sequence conditioned on the knowledge of the transmitted symbol sequence $p(\mathbf{y}_1^T \mid  \mathbf{x}_1^T)$, we write
\begin{equation}\label{eq:I_pyx}
    p(\mathbf{y}_1^T \mid  \mathbf{x}_1^T) = \frac{p(\mathbf{y}_1^T, \mathbf{x}_1^T)}{p(\mathbf{x}_1^T)} = \frac{\sum_{a_T \in {\mathcal{A}}}{p(a_T,\mathbf{y}_1^T, \mathbf{x}_1^T)}}{p(\mathbf{x}_1^T)}
\end{equation}
with the help of the auxiliary state. The computation of the joint probability $p(a_T,\mathbf{y}_1^T, \mathbf{x}_1^T)$ can be carried out efficiently using the forward recursion \eqref{eq:siso_alpha} with a modified branch matrix defined as
\begin{align}\label{eq:gamma_sdot_z}
    \gamma_{a,x}(a_t,a_{t-1}) &\triangleq p(y_t,x_t,a_t\mid a_{t-1}) \nonumber \\
    &=p(y_t \mid  x_t, a_t, a_{t-1}) \cdot  p(a_t \mid  a_{t-1}, x_t) \cdot p(x_t\mid a_{t-1}) \nonumber \\
    &=p(y_t \mid  a_t) \cdot  p(a_t \mid  a_{t-1}, x_t) \cdot p(x_t),
\end{align}
where the transition probability conditioned to the knowledge of transmitted symbols is written as
\begin{align}
    p(a_t& \mid  a_{t-1}, x_t =m\in\mathcal{X}) \nonumber \\  
    &= \begin{cases}
        P_{ij}, \quad &x_t = m, w_t=j, w_{t-1} = i, \\
        0, \quad & x_t \neq m,
    \end{cases}
\end{align}
where we use $m$ to denote a specific PSK symbol element in $\mathcal{X}$. Given \eqref{eq:I_py} and \eqref{eq:I_pyx}, the estimated AIR, represented by $\hat{I}(\mathbf{X};\mathbf{Y})$ can be computed via \eqref{eq:Iest}.

\section{System Design}\label{sec:receiver}
In Fig.~\ref{fig:block_tx}, we show the overall structure and components of the transmission system over a Markov-Middleton IN channel considered in this paper. This section will first introduce the bit-interleaved convolutional-coded DE-PSK transmitter and then detail the design methodology for the corresponding turbo-DE-PSK-IN receiver.

\subsection{Bit-Interleaved Convolutional-Coded DE-PSK Transmitter}\label{sec:tx}
The transmitter employs a serially concatenated bit-interleaved coded modulation scheme. In this scheme, a rate~$R$ convolutional encoder with $L$-stage shift register takes as input the information bit sequence $\mathbf{b}_1^K$ and generates coded bit sequence output $\mathbf{c}_1^{K/R}$, where $ b_k, c_k \in \{0,1\}$, based on a certain generator constraint. A bit-level interleaver ($\Pi$) is used to permute the convolutional encoder output, where the interleaved coded bit sequence is represented as $\mathbf{d}_1^{K/R}$. The bit-interleaver is designed to transform burst errors into single-bit errors that ensure that errors are distributed to maximize error correction capability, which is essential for turbo decoding. After the interleaver, a symbol sequence $\mathbf{x}_1^T$ is formed by a memoryless PSK symbol mapper, which maps a number of binary coded bits to a PSK symbol with the symbol time index denoted by the letter $t$. Finally, a differential symbol encoder modulates the incoming PSK symbols and produces 
\begin{align} \label{eq:diff_mod}
    z_t = x_t \cdot z_{t-1},    
\end{align}
with $z_0=1$, so the received signal is 
\begin{align}\label{eq:y_dqpskin}
    y_t =  x_t \cdot z_{t-1} + n_t = z_t + n_t.
\end{align}
This differential modulation introduces the recursive channel input–output relationship in \eqref{eq:y_dqpskin}.

Differential encoders are used in various wireless communication standards. One of the most significant advantages of a differential encoder is that it avoids the need for the receiver to have an exact phase reference, as the information is encoded in the phase difference of consecutive symbols. In this paper, we treat the differential encoder as a rate one convolutional code \cite{Narayanan99,Lee00} and thus create a serial-like turbo structure with the convolutional encoder. Its recursive structure largely enhances the time diversity in the serial turbo receiver design, leading to significant performance improvement compared to memoryless PSK modulation schemes \cite{Hoeher99,WvH12}. This study assumes a coherent detection scenario with a known symbol phase. As a result, we will ignore the reference symbols $z_0$ and $y_0$ in the subsequent analysis.

\begin{figure}[t]
    \centering
    \begin{tikzpicture}[line width=1.5pt, 
block1/.style={rectangle, fill=white, text opacity=1, rounded corners, draw, inner sep=2pt, minimum width=8mm, minimum height=8mm, align=center, font=\scriptsize, thick},
linestyle/.style = {-latex,>={Latex[length=1mm,width=1mm]},thick},]

\tikzstyle{block2} = [block1, fill=gray!40]
\tikzstyle{block3} = [block1, fill=LimeGreen]

\def\dist{2.5}
\def\ttsize{\scriptsize}

\node[] at (0,\dist) (b) {};
\node[block1, right=0.3*\dist of b, align=center] (Inner) {Conv.\\Enc.};
\draw[linestyle] ([xshift=-20pt]Inner.west) -- (Inner.west) node[midway,above] {\ttsize{$\mathbf{b}_1^K$}};

\node[block1, right=0.3*\dist of Inner.east, align=center] (PI) {$\Pi$};
\draw[linestyle] (Inner.east) -- (PI.west) node[midway,above] {\ttsize{$\mathbf{c}_1^{K/R}$}};

\node[block1, right=0.3*\dist of PI.east, align=center] (Map) {Sym.\\Map.};
\draw[linestyle] (PI.east) -- (Map.west) node[midway,above] {\ttsize{$\mathbf{d}_1^{K/R}$}};

\node[block1, right=0.3*\dist of Map.east, align=center] (DE) {Diff.\\Enc.};
\draw[linestyle] (Map.east) -- (DE.west) node[midway,above] {\ttsize{$\mathbf{x}_1^T$}};


\node[block1, right=0.3*\dist of DE.east, align=center] (IN) {IN};
\draw[linestyle] (DE.east) -- (IN.west) node[midway,above] {\ttsize{$\mathbf{z}_0^T$}};
\draw[linestyle] (IN.east) -- ([xshift=20pt]IN.east) node[midway, above] {\ttsize{$\mathbf{y}_0^T$}};


\end{tikzpicture}
    \caption{Block diagram of a convolutional-coded bit-interleaved DE-PSK transmission over an impulsive noise channel}
    \label{fig:block_tx}
\end{figure}
\begin{figure}[t]
\centering
    \centering
    \subfloat[Joint receiver design\label{fig:block_rx_joi}]{%
        \centering
        {\begin{tikzpicture}[line width=1.5pt, 
block1/.style={rectangle, fill=white, text opacity=1, rounded corners, draw, inner sep=2pt, minimum width=8mm, minimum height=8mm, align=center, font=\scriptsize,thick},
linestyle/.style = {-latex,>={Latex[length=1mm,width=1mm]},thick},]

\tikzstyle{block2} = [block1, fill=gray!40]
\tikzstyle{block3} = [block1, fill=LimeGreen]

\def\dist{2.5}
\def\ttsize{\scriptsize}

\node[] at (0,\dist) (b) {};
\node[block1, right=0*\dist of b, align=center] (det) {Joint DPSK-IN\\Demap.};
\draw[linestyle] ([xshift=-20pt]det.west) -- (det.west) node[midway,above] {\ttsize{$\mathbf{y}_1^T$}};

\node[block1, right=0.52*\dist of det.east, align=center] (PIInv) {\ttsize{$\Pi^{-1}$}};
\draw[linestyle] (det.east) -- (PIInv.west) node[midway,above] {\ttsize{$p^C_e(d_k)$}};

\node[block1, right=0.52*\dist of PIInv.east, align=center] (CC) {Conv.\\Dec.};
\draw[linestyle] (PIInv.east) -- (CC.west) node[midway,above] {\ttsize{$p^C_e(c_k)$}};
\draw[linestyle] (CC.east) -- ([xshift=20pt]CC.east) node[midway, above] {\ttsize{$\hat{\mathbf{b}}_1^K$}};

\node[block1, below=0.2*\dist of PIInv] (PI) {\ttsize{$\Pi$}};
\draw[linestyle] (CC.south) |- (PI.east) node[pos=0.8,above] {\ttsize{$p^I_e(c_k)$}};
\draw[linestyle] (PI.west) -| (det.south) node[pos=0.2,above] {\ttsize{$p^I_e(d_k)$}};


\end{tikzpicture}}     
        }
    \vspace{5mm} 
    \subfloat[Separate receiver design\label{fig:block_rx_sep}]{%
        \centering
        {\begin{tikzpicture}[line width=1.5pt, 
block1/.style={rectangle, fill=white, text opacity=1, rounded corners, draw, inner sep=2pt, minimum width=8mm, minimum height=8mm, align=center, font=\scriptsize,thick},
linestyle/.style = {-latex,>={Latex[length=1mm,width=1mm]},thick},]

\tikzstyle{block2} = [block1, fill=gray!40]
\tikzstyle{block3} = [block1, fill=LimeGreen]

\def\dist{2.5}
\def\ttsize{\scriptsize}

\node[] at (0,\dist) (b) {};
\node[block1, right=0*\dist of b, align=center] (det) {IN\\Det.};
\draw[linestyle] ([xshift=-20pt]det.west) -- (det.west) node[midway,above] {\ttsize{$\mathbf{y}_1^T$}};

\node[block1, right=0.52*\dist of det.east, align=center] (demap) {DPSK\\Demap.};
\draw[linestyle] (det.east) -- (demap.west) node[midway,above] {\ttsize{$p(z_t,\mathbf{y}_1^T)$}};

\node[block1, right=0.52*\dist of demap.east, align=center] (PIInv) {\ttsize{$\Pi^{-1}$}};
\draw[linestyle] (demap.east) -- (PIInv.west) node[midway,above] {\ttsize{$p^C_e(d_k)$}};

\node[block1, right=0.52*\dist of PIInv.east, align=center] (CC) {Conv.\\Dec.};
\draw[linestyle] (PIInv.east) -- (CC.west) node[midway,above] {\ttsize{$p^C_e(c_k)$}};
\draw[linestyle] (CC.east) -- ([xshift=20pt]CC.east) node[midway, above] {\ttsize{$\hat{\mathbf{b}}_1^K$}};

\node[block1, below=0.2*\dist of PIInv] (PI) {\ttsize{$\Pi$}};
\draw[linestyle] (CC.south) |- (PI.east) node[pos=0.8,above] {\ttsize{$ p^I_e(c_k)$}};
\draw[linestyle] (PI.west) -| (demap.south) node[pos=0.2,above] {\ttsize{$p^I_e(d_k)$}};


\end{tikzpicture}}     
        }
\caption{Block diagrams of (a) optimal joint and (b) the suboptimal separate turbo-DE-PSK-IN receiver designs.}
\end{figure}

\subsection{Optimal Joint Turbo-DE-PSK-IN Receiver}\label{sec:rec_joi}
\subsubsection{MAP DE-PSK-IN demapper}
In the joint detector-demapper design, we construct a super-trellis where the structure of the Markov-Middelton IN detector and the DE-PSK demapper are considered jointly. We define the state realization of the joint super-trellis as 
\begin{align}\label{eq:state_dpskin}
     \dot{a}_t \triangleq f(z_t\in \mathcal{X},w_t\in \mathcal{W}) \in \mathcal{A},
\end{align}
which is drawn from the same state space as $a_t$ in \eqref{eq:state_qpskin} and has a similar form of transition probability 
\begin{align} \label{eq:P_n_x_joint}
    p(\dot{a}_t \mid  &\dot{a}_{t-1}) = P_{ij} \cdot p(x_t), \quad w_t=j, w_{t-1} = i.
\end{align}
However, instead of the PSK symbol $x_t$, the state $\dot{a}_t$ is now associated with the differentially encoded symbol $z_t$ with a transition driven by $x_t$. The transition of the joint super-trellis section for a DE-QPSK modulation with a $2$-state Markov-Middleton model is visualized in Fig.~\ref{fig:trellis_IN2DQPSK}. The likelihood function, on the other hand, can be written as
\begin{align}\label{eq:lik_joint}
	p(y_t\mid \dot{a}_t) =\frac{1} {{\pi \sigma^2_{j}}}\exp\left(-\frac{|y_{t}-z_t|^2}{\sigma^2_{j}}\right), \quad w_t = j.
\end{align}

We can now define the branch metric of the joint DE-PSK-IN detector-demapper as 
\begin{align}\label{eq:gamma_adot}
    \gamma_{\dot{a}}(\dot{a}_t,\dot{a}_{t-1}) &\triangleq  p(y_t\mid \dot{a}_t) \cdot p(\dot{a}_t\mid \dot{a}_{t-1}).
\end{align}
Given the transition probability \eqref{eq:P_n_x_joint} and the likelihood function \eqref{eq:lik_joint} for computing the branch metric \eqref{eq:gamma_adot}, the joint probability $p(\dot{a}_t, \dot{a}_{t-1}, \mathbf{y}_1^T)$ of the joint DE-PSK-IN detector-demapper is computed via \eqref{eq:bcjr}--\eqref{eq:siso_beta}. A MAP-based soft-output detection is then performed through 
\begin{align}
     p(x_t, \mathbf{y}_1^T) = \sum_{(\dot{a}_{t-1}\rightarrow \dot{a}_t): x_t}{ {p( \dot{a}_t, \dot{a}_{t-1}, \mathbf{y}_1^T)} }. \nonumber
\end{align}
where $(\dot{a}_{t-1} \rightarrow \dot{a}_t):x_t$ denotes all the state transitions $(\dot{a}_{t-1} \rightarrow \dot{a}_t)$ that is driven by the input symbol $x_t$.

\subsubsection{Iterative Turbo Decoder}\label{sec:it_dec}
To perform bit deinterleaving, we need to convert the symbol probability to bit probability as
\begin{align} \label{eq:turbo_s2b}
    p(d_k, \mathbf{y}_1^T) = \Phi^{-1}[p(x_t,\mathbf{y}_1^T)]
\end{align}
and 
\begin{align}
    p^I_e(d_k) = \Phi^{-1}[p^I_e(x_t)], \nonumber  
\end{align}
where $\Phi^{-1}$ denotes the inverse symbol mapping function that converts the probability of the PSK symbol $x_t$ back to the probability of the corresponding bit $d_k$.

One essential part of the turbo decoding principle is to pass only the extrinsic information to each MAP decoder (demapper) block. Extrinsic information refers to the information provided by one decoder (demapper) about one specific bit (symbol), excluding the information directly used to derive it. In other words, it represents new information that can help the other decoder (demapper) refine its estimates. Here, we define the extrinsic likelihood information from the joint DE-PSK-IN demapper as the joint probability $p(d_{k}, \mathbf{y}_1^T)$ divided by the prior information regarding $d_k$ as
\begin{align}
     p^C_e(d_{k}) \triangleq \frac{p(d_{k}, \mathbf{y}_1^T)}{p^I_e(d_{k})}. \nonumber
\end{align}
After performing the bit-deinterleaving, we can derive the extrinsic likelihood information of the coded bits as 
\begin{align}
     p^C_e(c_{k}) = \Pi^{-1} \left[ p^C_e(d_k) \right], \nonumber
\end{align}
which is then fed to the convolutional decoder, whose state realization is defined as 
\begin{align}
    g_k &\triangleq f(b_k,b_{k-1},\cdots,b_{k-L-1} \in \mathcal{U}^L) \in \mathcal{G} \triangleq \{0,1,\cdots, 2^L-1\}, \nonumber
\end{align}
where $f:\mathcal{U}^L \rightarrow \mathcal{G}$ maps $\mathcal{U}^L$ to distinct states in $\mathcal{G}$. The modified branch metric for the convolutional decoder is then written as 
\begin{align}\label{eq:gamma_conv}
    \gamma_g(g_k, g_{k-1}) \triangleq p^C_e(c_k) \cdot  p(b_k), \quad (g_{k-1}\rightarrow g_k): b_k. 
\end{align}
The extrinsic likelihood information  $p^C_e(c_k)$ produced by the DE-PSK demapper acts as the convolutional decoder's likelihood information, and $p(b_k)=1/2$ is the prior bit probability. Again, we compute the joint probability $p(g_k,g_{k-1}, \mathbf{y}_1^T)$ via \eqref{eq:bcjr}--\eqref{eq:siso_beta} using \eqref{eq:gamma_conv} and obtain the joint probability of the information bit as
\begin{align} 
    p(b_k, \mathbf{y}_1^T) =  {\sum_{(g_{k-1}\rightarrow g_k): b_k}{{ p(g_k,g_{k-1}, \mathbf{y}_1^T)} }}, \nonumber
\end{align}
and the joint probability of the coded bit as
\begin{align}
    p(c_k, \mathbf{y}_1^T) =  {\sum_{(g_{k-1}\rightarrow g_k) : c_k}{{p(g_k, g_{k-1}, \mathbf{y}_1^T)} }}. \nonumber
\end{align}
To compute the extrinsic prior information produced by the convolutional decoder, we divide the joint probability \( p(c_k, \mathbf{y}_1^T) \) by the likelihood function \( p^C_e(c_k) \) used for deriving it as
\begin{align}
    p^I_e(c_k) \triangleq \frac{p(c_k,\mathbf{y}_1^T)}{p^C_e(c_k)}.  \nonumber
\end{align}
After the bit interleaver 
\begin{align}
    p^I_e(d_k) = \Pi \left[ p^I_e(c_k) \right],  \nonumber
\end{align}
we can derive the extrinsic prior information of the PSK symbol via bit-to-symbol mapping  
\begin{align}
    p^I_e(x_t) = \Phi[p^I_e(d_{k})]. \nonumber
\end{align}
The extrinsic symbol information $p^I_e(x_t)$ generated by the convolutional decoder is then fed back as the prior information in \eqref{eq:turbo_gamma1} of the joint DE-PSK-IN demapper for the next turbo iteration.

\subsection{Suboptimal Separate Turbo-DE-PSK-IN Receiver} \label{sec:rec_sep}
While the joint design delivers the optimal MAP solution by iteratively updating the extrinsic information for both the IN detector and DE demapper simultaneously, the computational complexity increases significantly with the number of turbo iterations. Therefore, we introduce in this section a suboptimal design that separates the IN detector from the turbo-DE-PSK demapper-decoder to reduce the operational complexity.

\subsubsection{MAP IN Detector}
The states of the standalone MAP-based IN detector are defined as in \eqref{eq:state_qpskin}, and a trellis section example is shown in Fig.~\ref{fig:trellis_IN2QPSK}. Given the transition probability \eqref{eq:P_dotstran}, likelihood function \eqref{eq:P_dotslik}, and the branch metric \eqref{eq:gamma_sdot}, the joint probability $p(a_t, a_{t-1}, \mathbf{y}_1^T)$ of the IN detector is computed via \eqref{eq:bcjr}--\eqref{eq:siso_beta}. A MAP-based soft-output detection is then performed through 
\begin{align} \label{eq:siso_det_z}
     p(z_t, \mathbf{y}_1^T) = \sum_{(a_{t-1}\rightarrow a_t): z_t}{ {p( a_t, a_{t-1}, \mathbf{y}_1^T)} }.
\end{align}
where $(a_{t-1} \rightarrow a_t):z_t$ denotes all the state transitions $(a_{t-1} \rightarrow a_t)$ that is driven by the input differentially encoded symbol $z_t$.

\subsubsection{MAP DE Demapper}
The joint probability \eqref{eq:siso_det_z} generated from the IN detector is used as a soft input, representing the channel information, to the differential symbol demapper. We define the state realization of the differential decoder as
\begin{align}\label{eq:state_dpskmap}
     l_t \triangleq f(z_t\in \mathcal{X}) \in \mathcal{L}  \triangleq \{0,1,\cdots,M-1\},
\end{align}
where $f:\mathcal{X} \rightarrow \mathcal{L}$ maps $z_t \in \mathcal{X}$ to a specific state in $\mathcal{L}$. The trellis section of a DE-QPSK decoder is shown in Fig.~\ref{fig:tr_dqpsk}. Note that although the state realization $l_t$ is a mapping of the DE-PSK symbol $z_t$, the trellis input is the PSK modulated symbol $x_t$, which reveals the structure of the differential encoder in \eqref{eq:diff_mod}. 

Using the state definitions \eqref{eq:state_dpskmap} and the joint probability $p(z_t,\mathbf{y}_1^T)$ provided the IN detector, we can now rewrite \eqref{eq:gamma} and express the trellis branch metric for the differential decoder as 
\begin{align} \label{eq:turbo_gamma1}
    \gamma_p(l_t, l_{t-1}) \triangleq p(z_t,\mathbf{y}_1^T) \cdot p^I_e(x_t) , \quad (l_{t-1} \rightarrow l_t): x_t.
\end{align}
where the modified branch metric uses $p(z_t,\mathbf{y}_1^T)$ as the likelihood information and uses the extrinsic prior symbol information $p^I_e(x_t)$ to denote the transition probability $p(l_t\mid l_{t-1})$ when the state transition $(l_{t-1} \rightarrow l_t)$ is driven by the input symbol $x_t$. Note that we set  $p^I_e(x_t) = p(x_t) = 1/M$ in the first iteration. Using the modified branch metric \eqref{eq:turbo_gamma1}, the joint probability of the DE-PSK demapper $p(l_t, l_{t-1}, \mathbf{y}_1^T)$ is derived through \eqref{eq:bcjr}--\eqref{eq:siso_beta}, and the joint probability of the PSK symbol is then obtained by 
\begin{align}
    p(x_t, \mathbf{y}_1^T) = \sum_{(l_{t-1}\rightarrow l_t): x_t}{ {p(l_t,l_{t-1}, \mathbf{y}_1^T)} }. \nonumber
\end{align}
Subsequently, the iterative processing is the same as in Sec.~\ref{sec:it_dec}.

\begin{figure}[t]
    \centering
    {
\tikzstyle{state}=[shape=circle,draw=black!50,fill=gray!50, font=\small]
\tikzstyle{observation}=[shape=rectangle,draw=orange!50,fill=orange!20]
\tikzstyle{mainstate}=[state,very thick,]
\tikzstyle{edge1}=[stealth-, thick, Black]
\tikzstyle{edge2}=[stealth-, thick, Cyan]
\tikzstyle{edge3}=[stealth-, thick, Orange]
\tikzstyle{edge4}=[stealth-, thick, Green]
\begin{tikzpicture}[]

\def\dist{3}  
\def\v_dist{.8}  

\node(s')   at (0.3, 5.8*\v_dist) {$l_{t-1}$};
\node(s1)   at (-.6, 5*\v_dist) {\scriptsize$f(e^{j0})$};
\node       at (-.6, 4*\v_dist) {\scriptsize$f(e^{j\frac{\pi}{2}})$};
\node       at (-.6, 3*\v_dist) {\scriptsize$f(e^{j\pi})$};
\node       at (-.6, 2*\v_dist) {\scriptsize$f(e^{j\frac{3\pi}{2}})$};

\node[mainstate] (s1_1) at (0.3,5*\v_dist) {{$0$}};
\node[mainstate] (s2_1) at (0.3,4*\v_dist) {{$1$}};
\node[mainstate] (s3_1) at (0.3,3*\v_dist) {{$2$}};
\node[mainstate] (s4_1) at (0.3,2*\v_dist) {{$3$}};

\node     [right=\dist of s'] {$l_t$};
\node[mainstate] (s1_2) [right=\dist of s1_1] {$0$}
    edge[edge1] (s1_1)
    edge[edge4] (s2_1)
    edge[edge3] (s3_1)
    edge[edge2] (s4_1);
\node[mainstate] (s2_2) [right=\dist of s2_1] {$1$}
     edge[edge2] (s1_1)
     edge[edge1] (s2_1)
     edge[edge4] (s3_1)
     edge[edge3] (s4_1);
\node[mainstate] (s3_2) [right=\dist of s3_1] {$2$}
     edge[edge3] (s1_1)
     edge[edge2] (s2_1)
     edge[edge1] (s3_1)
     edge[edge4] (s4_1);
\node[mainstate] (s4_2) [right=\dist of s4_1] {$3$}
     edge[edge4] (s1_1)
     edge[edge3] (s2_1)
     edge[edge2] (s3_1)
     edge[edge1] (s4_1);

\node[
    anchor=west,
    fill=white,
    inner sep=0.1pt,
    font=\footnotesize
    ] at (s2_2.east) {
    \begin{tabular}{c@{} c@{} c@{} }
        \multicolumn{2}{c}{Trellis Input} \\ 
        \begin{tikzpicture}
            \draw[edge1, -stealth] (1, 0) -- (1.6,0);
        \end{tikzpicture} & \raisebox{-0.3ex}{$e^{j0}$} \\
        \begin{tikzpicture}
            \draw[edge2, -stealth] (1, 0) -- (1.6,0);
        \end{tikzpicture} & \raisebox{-0.3ex}{$e^{j\frac{\pi}{2}}$} \\
        \begin{tikzpicture}
            \draw[edge3, -stealth] (1, 0) -- (1.6,0);
        \end{tikzpicture} & \raisebox{-0.3ex}{$e^{j\pi}$} \\
        \begin{tikzpicture}
            \draw[edge4, -stealth] (1, 0) -- (1.6,0);
        \end{tikzpicture} & \raisebox{-0.3ex}{$e^{j\frac{3\pi}{2}}$} \\
    \end{tabular}
};

\end{tikzpicture}}     
    \caption{Trellis section of a standalone DE-QPSK demapper.}
    \label{fig:tr_dqpsk}
\end{figure}

\begin{table}[t]
\centering
\renewcommand{\arraystretch}{1.3}
\caption{Number of multiplications for separate and joint turbo receiver designs with $M=4$~,~$W=4$~, and~$L=2$.}
    \begin{tabular}{|c|c|c|c|c|c|}
    \hline
        & $I=0$   & $I=1$  & $I=2$ & $I=3$ & $I=10$  \\ \hline
    $\mathcal{C}_\text{joi}$ & $1024T$   & $2048T$ & $3072T$ & $4096T$ & $11264T$ \\ \hline
    $\mathcal{C}_\text{sep}$ & $1056T$   & $1600T$ & $2144T$ & $2688T$ & $6496T$ \\ \hline
    \end{tabular}
\label{tab:complexity}
\end{table}
\subsection{Computational Complexity Analysis}
This subsection offers a general comparison of the complexity between the joint and separate system designs. Since the main computational burden comes from the multiplication operation in the BCJR algorithm, we will concentrate on the complexity related to the number of multiplications used in the BCJR algorithm for two different realizations of the turbo systems.

Consider a length $T$ sequence ($T$ trellis sections) processed by a BCJR algorithm with $N$ trellis states. For each state, it needs to consider transitions from all possible previous states, which results in a $N^2$ multiplication per trellis section. Therefore, the overall complexity of a BCJR operation is given by $2T N^2$, where the constant $2$ accounts for the forward and backward operations. For transmission of a general cardinality $M$ and $W$ state Markov-Middleton channel, the super-trellis sections of the joint DE-PSK-IN demapper (Sec.~\ref{sec:rec_joi}) require $MW$ states. Considering the turbo receiver operates over $I$ iterations, we can then write the computational complexity of the joint receiver as
\begin{align}
    \mathcal{C}_\text{joi} = 2T \cdot (I+1)( M^2W^2 + M^{2L} ). \nonumber
\end{align}
The separate design (Sec.~\ref{sec:rec_sep}), on the other hand, requires an $MW$ state MAP IN detector, an $M$ state MAP DE-PSK demapper, and an $M^L$ state MAP convolutional decoder. We can then write its computational complexity as 
\begin{align}
    \mathcal{C}_\text{sep} = 2T \cdot ( M^2 W^2 + (I+1)(M^2+M^{2L}) ). \nonumber
\end{align}

In Table~\ref{tab:complexity}, we show the computational complexity of $\mathcal{C}_\text{joi}$ and $\mathcal{C}_\text{sep}$ for a transmission of a $L=2$ convolutional code and a DE-QPSK ($M=4$) modulation over a $4$-state ($W=4$) Markov Middleton model. When operating without turbo iterations ($I=1$), the super-trellis design in the joint system has a slightly lower complexity than a separate design. As the number of iterations increases, the computational cost of the joint design increases significantly and exceeds that of a separate design from the second turbo iteration. After $10$ iterations, the suboptimal separate design has nearly half the computational complexity of the optimal receiver design.

\section{Simulation Results}\label{sec:simu}
\begin{figure}[t]
    \centering
    \resizebox{!}{0.5\textwidth}{	
\begin{tikzpicture}
    \begin{semilogxaxis}[
        axis line style = thick,
        grid = both,
        name = p1,
        xmin = 10^-2, xmax=10^4,
        ymin = 0.58, ymax=1.5,
        font=\footnotesize,
        xlabel = $\Lambda$, ylabel = AIR (bits/symbol),
        legend style={
            font=\small,
            nodes={scale=1.0},
        },
        legend cell align={left},
        legend pos = north east,
        ]
        \addplot[very thick, color=Gray] 
        coordinates {(10^-2, 1.44) (10^5, 1.44)};
        \draw[->, thick, color=Gray] 
            (axis cs: 2*10^-1, 1.44) 
                -- (axis cs: 10^0, 1.38); 
            \node at (axis cs: 10^0, 1.38) [right] {\footnotesize{QPSK-AWGN ($\text{AIR}=1.44$)}};
            

        \addplot[color=Red, very thick, mark=+, mark size=6pt] coordinates {(0.1,1.38)};
        
        \addplot[color=Orange, very thick, dotted] 
        table[x expr=\thisrowno{0}, y expr=\thisrowno{1}] 
        {AIRQPSK_SNR3.txt}; \label{air_r0a1} 
        
        \addplot[color=RoyalBlue,  very thick, dotted] 
        table[x expr=\thisrowno{0}, y expr=\thisrowno{2}] 
        {AIRQPSK_SNR3.txt}; \label{air_r5a1}
    
        \addplot[color=Black, very thick, dotted] 
        table[x expr=\thisrowno{0}, y expr=\thisrowno{3}] 
        {AIRQPSK_SNR3.txt}; \label{air_r9a1}
        
        \addplot[color=Orange, very thick,dashed] 
        table[x expr=\thisrowno{0}, y expr=\thisrowno{4}] 
        {AIRQPSK_SNR3.txt}; \label{air_r0a3} 
        
        \addplot[color=RoyalBlue,  very thick,dashed] 
        table[x expr=\thisrowno{0}, y expr=\thisrowno{5}] 
        {AIRQPSK_SNR3.txt}; \label{air_r5a3}
    
        \addplot[color=Black, very thick, dashed] 
        table[x expr=\thisrowno{0}, y expr=\thisrowno{6}] 
        {AIRQPSK_SNR3.txt}; \label{air_r9a3}

        \addplot[color=Orange, very thick] 
        table[x expr=\thisrowno{0}, y expr=\thisrowno{7}] 
        {AIRQPSK_SNR3.txt}; \label{air_r0a5} 
        
        \addplot[color=RoyalBlue,  very thick] 
        table[x expr=\thisrowno{0}, y expr=\thisrowno{8}] 
        {AIRQPSK_SNR3.txt}; \label{air_r5a5}
    
        \addplot[color=Black, very thick] 
        table[x expr=\thisrowno{0}, y expr=\thisrowno{9}] 
        {AIRQPSK_SNR3.txt}; \label{air_r9a5}
        
    \end{semilogxaxis}

    \matrix[
    matrix of nodes,
    anchor=south west,
    fill = white,draw,
    inner sep = 0.1em,
    column sep = 0.1em,
    node font=\scriptsize,
      ]
      at ([xshift=2pt, yshift=2pt]current axis.south west){
        $r \backslash A$  & $0.1$   & $0.3$ & $0.5$\\ 
        $0$   & \ref{air_r0a1} & \ref{air_r0a3} & \ref{air_r0a5} \\
        $0.5$ & \ref{air_r5a1} & \ref{air_r5a3} & \ref{air_r5a5} \\
        $0.9$ & \ref{air_r9a1} & \ref{air_r9a3} & \ref{air_r9a5} \\
    };
\end{tikzpicture}    }
    \caption{AIRs versus impulsive-to-background noise ratio $\Lambda$ for different impulsive index $A=0.1\text{, }0.3\text{, and } 0.5$ and correlation parameter $r=0\text{, }0.5\text{, and } 0.9$ for a QPSK modulation scheme with a $4$-state Markov-Middleton impulsive noise model with SNR $=3$~dB.}
    \label{fig:airG}
\end{figure}

\begin{figure}[t]
    \centering
    \subfloat[$\Lambda=10^{-2}$\label{fig:probG2}]{
        \resizebox{0.33\linewidth}{!}{\begin{tikzpicture}
	
	\begin{axis}[
		axis line style = thick,
		grid = both,
		name = p1,
		xmin = -3, xmax=5,
		ymin = -15, ymax=0,
		xtick = {-3, -1, 1, 3, 5},{font=\scriptsize},
		font=\small,
		ylabel = $\ln{p(y_t|x_t)} \text{,} \ln{p(y_t|x_t,w_t)}$, xlabel = $x_t$,
        ylabel style={yshift=-4pt},
		legend style={
			font=\tiny,
			nodes={scale=1.0},
		},
		legend cell align={left},
		legend pos = south west,
		]

            
		  \addplot[color=Black, very thick] 
             table[x expr=\thisrowno{0}, y expr=\thisrowno{1}] 
            {Prob_pyx_G2.txt}; 


            \addplot[color=Black, thick, dotted, mark=*, mark options={solid,fill=white}, mark repeat=3, mark size=2.5] 
             table[x expr=\thisrowno{0}, y expr=\thisrowno{3}] 
            {Prob_pyx_G2.txt}; 
            
            \addplot[color=Black, thick, dotted, mark=asterisk, mark options={solid,fill=white}, mark repeat=3, mark size=2.5] 
             table[x expr=\thisrowno{0}, y expr=\thisrowno{4}] 
            {Prob_pyx_G2.txt}; 

            


	\end{axis}

\end{tikzpicture}    }
    }
    \subfloat[$\Lambda=10^{0}$\label{fig:probG0}]{
        \resizebox{0.33\linewidth}{!}{\begin{tikzpicture}
	
	\begin{axis}[
		axis line style = thick,
        legend to name=commonlegend,
		grid = both,
		name = p1,
		xmin = -3, xmax=5,
		ymin = -15, ymax=0,
		xtick = {-3, -1, 1, 3, 5},
		font=\small,
		ylabel = $\ln{p(y_t|x_t)} \text{,} \ln{p(y_t|x_t,w_t)}$, xlabel = $x_t$,
        ylabel style={yshift=-4pt},
		legend style={
			font=\small,
			nodes={scale=1.0},
                legend columns = 3,
                column sep=2pt
		},
		legend cell align={left},
		legend pos = south west,
		]
            
		  \addplot[color=Black, very thick] 
             table[x expr=\thisrowno{0}, y expr=\thisrowno{1}] 
            {Prob_pyx_G0.txt}; \addlegendentry{$p(y_t|x_t=+1)$};

            \addplot[color=Black, thick, dotted, mark=*, mark options={solid,fill=white}, mark repeat=3, mark size=2.5] 
             table[x expr=\thisrowno{0}, y expr=\thisrowno{3}] 
            {Prob_pyx_G0.txt}; \addlegendentry{$p(y_t|x_t=+1,w_t=0)$};

           \addplot[color=Black, thick, dotted, mark=asterisk, mark options={solid,fill=white}, mark repeat=6, mark size=2.5] 
             table[x expr=\thisrowno{0}, y expr=\thisrowno{4}] 
            {Prob_pyx_G0.txt}; \addlegendentry{$p(y_t|x_t=+1,w_t=1)$};	

	\end{axis}

\end{tikzpicture}    }
    }
    \subfloat[$\Lambda=10^{4}$\label{fig:probG-4}]{
        \resizebox{0.33\linewidth}{!}{\begin{tikzpicture}
	
	\begin{axis}[
		axis line style = thick,
		grid = both,
		name = p1,
		xmin = -3, xmax=5,
		ymin = -15, ymax=0,
		xtick = {-3, -1, 1, 3, 5},
		font=\small,
		ylabel = $\ln{p(y_t|x_t)} \text{,} \ln{p(y_t|x_t,w_t)}$, xlabel = $x_t$,
        ylabel style={yshift=-4pt},
		legend style={
			font=\tiny,
			nodes={scale=1.0},
		},
		legend cell align={left},
		legend pos = south west,
		]
            
		  \addplot[color=Black, very thick] 
             table[x expr=\thisrowno{0}, y expr=\thisrowno{1}] 
            {Prob_pyx_G-4.txt}; 
            

            \addplot[color=Black, thick, dotted, mark=*, mark options={solid,fill=white}, mark repeat=3, mark size=2.5] 
             table[x expr=\thisrowno{0}, y expr=\thisrowno{3}] 
            {Prob_pyx_G-4.txt}; 
            
            \addplot[color=Black, thick, dotted,  mark=asterisk, mark options={solid,fill=white}, mark repeat=10, mark size=2.5] 
             table[x expr=\thisrowno{0}, y expr=\thisrowno{4}] 
            {Prob_pyx_G-4.txt}; 

            


	\end{axis}

\end{tikzpicture}    }
    }
    \vspace{1ex}
    \pgfplotslegendfromname{commonlegend}

    \caption{The likelihood functions $p(y_t|x_t=+1)$, $p(y_t|x_t=+1,w_t=0)$, and $p(y_t|x_t=+1,w_t=1)$ of a 2-state memoryless ($r=0$) Markov--Middleton model with BPSK modulation are shown for (a) $\Lambda=10^{-2}$, (b) $\Lambda=10^0$, and (c) $\Lambda=10^{4}$.}
    \label{fig:prob}
\end{figure}
\begin{figure}[t]
    \centering
     \resizebox{!}{0.5\textwidth}{	
\begin{tikzpicture}
    \begin{semilogxaxis}[
        axis line style = thick,
        grid = both,
        name = p1,
        xmin = 10^-2, xmax=10^4,
        ymin = 0.88, ymax=1.45,
        font=\footnotesize,
        xlabel = $\Lambda$, ylabel = AIR (bits/symbol),
        legend style={
            font=\tiny,
            nodes={scale=1.0},
            legend columns=2, 
            reverse legend,   
        },
        legend cell align={left},
        legend pos = north east,
        ]

        \addplot[color=RoyalBlue, thick] 
        table[x expr=\thisrowno{0}, y expr=\thisrowno{4}] 
        {AIRQPSKmis_SNR3.txt}; \label{misair_a1};\addlegendentry{$\hat{A}=0.1$}
        
        \addplot[color=RoyalBlue, thick, dashed] 
        table[x expr=\thisrowno{0}, y expr=\thisrowno{5}] 
        {AIRQPSKmis_SNR3.txt}; \label{misair_a01};\addlegendentry{$\hat{A}=0.01$}

        \addplot[color=ForestGreen, thick] 
        table[x expr=\thisrowno{0}, y expr=\thisrowno{6}] 
        {AIRQPSKmis_SNR3.txt}; \label{misair_r5};\addlegendentry{$\hat{r}=0.5$};
        
        \addplot[color=ForestGreen, thick, dashed] 
        table[x expr=\thisrowno{0}, y expr=\thisrowno{7}] 
        {AIRQPSKmis_SNR3.txt}; \label{misair_r0};\addlegendentry{$\hat{r}=0$};

        \addplot[color=Orange, thick] 
        table[x expr=\thisrowno{0}, y expr=\thisrowno{9}] 
        {AIRQPSKmis_SNR3.txt}; \label{misair_l01};\addlegendentry{$\hat{\Lambda}=10^{-2}$};
        
        \addplot[color=Orange, thick, dashed] 
        table[x expr=\thisrowno{0}, y expr=\thisrowno{8}] 
        {AIRQPSKmis_SNR3.txt}; \label{misair_l100};\addlegendentry{$\hat{\Lambda} = 10^2$};


        \addplot[color=Black, thick, mark=x, mark size=2.2,mark repeat=2] 
        table[x expr=\thisrowno{0}, y expr=\thisrowno{2}] 
        {AIRQPSKmis_SNR3.txt}; \label{misair_n2};\addlegendentry{$\hat{W}=2$};

        \addplot[color=Black, thick, dashed] 
        table[x expr=\thisrowno{0}, y expr=\thisrowno{3}] 
        {AIRQPSKmis_SNR3.txt}; \label{misair_n1};\addlegendentry{$\hat{W}=1$};

        \addplot[color=Black, thick, mark=o, mark options={solid,fill=white}, mark size=2.8, mark repeat=2] 
        table[x expr=\thisrowno{0}, y expr=\thisrowno{6}] 
        {AIRQPSK_SNR3.txt};\addlegendentry{REF};

    \end{semilogxaxis}

\end{tikzpicture}    }
    \caption{AIRs versus impulsive-to-background noise ratio $\Lambda$ for mismached $\hat{A}=0.1\text{ and }0.01$, $\hat{r}=0\text{ and } 0.5$, $\hat{\Lambda}=10^{-2} \text{ and } 10^2$, and $\hat{W}=1\text{ and }2$, for a QPSK modulation scheme with SNR $=3$~dB. The AIR with actual channel parameters $A=0.3 \text{, } r=0.9 \text{, and } W=4$ is denoted as REF.}
    \label{fig:airG_mis}
\end{figure}
\begin{figure}[t]
    \centering
    \resizebox{!}{0.5\textwidth}{	
\begin{tikzpicture}
    \begin{axis}[
        axis line style = thick,
        grid = both,
        name = p1,
        xmin = -6, xmax=6,
        ymin = 0.1, ymax=1.9,
        font=\footnotesize,
        xlabel = SNR (dB), ylabel = AIR (bits/symbol),
        legend style={
            font=\footnotesize,
            nodes={scale=1.0},
        },
        legend cell align={left},
        legend pos = north west,
        ]
        \addplot[color=Gray, very thick] 
        table[x expr=\thisrowno{0}, y expr=\thisrowno{1}] 
        {AIRQPSK_SNR.txt}; \label{airsnr_awgn}; \addlegendentry{QPSK-AWGN};

        \addplot[color=Black, dotted, very thick] 
        table[x expr=\thisrowno{0}, y expr=\thisrowno{2}] 
        {AIRQPSK_SNR.txt}; \label{airsnr_A01}; \addlegendentry{$A=0.1$};

        \addplot[color=Black, dashed, very thick] 
        table[x expr=\thisrowno{0}, y expr=\thisrowno{3}] 
        {AIRQPSK_SNR.txt}; \label{airsnr_A03}; \addlegendentry{$A=0.3$};

        \addplot[color=Black, very thick] 
        table[x expr=\thisrowno{0}, y expr=\thisrowno{4}] 
        {AIRQPSK_SNR.txt}; \label{airsnr_A05}; \addlegendentry{$A=0.5$};

        \addplot[very thick, color=Red, dotted] 
        coordinates {(-10, 1) (4.2, 1)};
        \addplot[very thick, color=Red, dotted] 
        coordinates {(4.2, 1) (4.2, -0.1)};
        \addplot[very thick, color=Red, dotted] 
        coordinates {(2.4, 1) (2.4, -0.1)};
        \addplot[very thick, color=Red, dotted] 
        coordinates {(0.9, 1) (0.9, -0.1)};

        \node at (axis cs: 1.2, .5) [left, rotate=90]  {\footnotesize{\textcolor{red}{0.9~dB}}};
        \node at (axis cs: 2.7, .5) [left, rotate=90]  {\footnotesize{\textcolor{red}{2.4~dB}}};
        \node at (axis cs: 4.5, .5) [left, rotate=90]  {\footnotesize{\textcolor{red}{4.2~dB}}};
    \end{axis}

\end{tikzpicture}    }
    \caption{AIRs versus SNR for QPSK transmission over Markov-Middleton IN channels for $A=0.1 \text{, }0.3 \text{, and } 0.5$ with fixed $\Lambda=10$, $r=0.9$, and $W=4$. The solid gray line indicates the AWGN channel.}
    \label{fig:air_snr}
\end{figure}

\subsection{AIR Analysis}\label{sec:simu_air}
In this section, we report a detailed AIR analysis for the finite state Markov-Middleton model with a PSK transmission scheme. Specifically, we simulated $1000$ independent sequences with a length of $T=10^6$ symbols to compute the estimated AIR, where the transmitted sequence $\mathbf{x}_1^T$ consists of QPSK ($M=4$) modulated symbols with a symbol set $\mathcal{X}$ following \eqref{eq:xsym}. The channel output sequence $\mathbf{y}_1^T$ is obtained using \eqref{eq:airio}, where the noise sequence $\mathbf{n}_1^T$ is generated by a $4$-state Markov-Middleton model with noise statistics following \eqref{eq:MMApdf2}--\eqref{eq:varm}, given parameters $A$, $\Lambda$, $r$, and $\sigma_0^2$. The choice of $W=4$ is based on both empirical evidence from our field-testing data \cite{chen2024} and a balanced trade-off between model accuracy and computational complexity. Note that although the mathematical derivations of this paper are in the probability domain, the actual implementation is in the log domain to prevent numerical instability. The simulated AIR with different Markov-Middleton channel parameters are shown in Fig.~\ref{fig:airG} for an SNR of $3$~dB. Here, the SNR is defined as the transmitted symbol power over the background noise power 
\begin{align}
    \text{SNR}=E\{|x_t|^2\}/\sigma_0^2 \nonumber.
\end{align}
To help illustrate the behavior of the AIR over impulsive-to-background noise power ratio $\Lambda$, we also plot the likelihood function for a memoryless ($r=0$) channel with different impulsive-to-background noise power ratios ($\Lambda=10^{-2},10^0,10^4$) in Figs.~\ref{fig:probG2}--\ref{fig:probG-4}. For simplicity, we choose a two-state (background and impulsive states) channel with impulsive index $A=0.3$ and BPSK for the likelihood function analysis, which does not affect the generality of the presentation.

Results from Fig.~\ref{fig:airG} show that a BN-limited regime can be found when $\Lambda < 10^{-1}$ (indicated by red cross), where the noise generated by an IN state is barely distinguishable from the noise generated by a BN state ($\sigma^2_0 \approx \sigma^2_j, \quad j=1,2,3$) as shown in the likelihood function from Fig.~\ref{fig:probG2}. In this case, the performance of the Markov-Middleton channel approaches that of the AWGN channel, with an AIR of $1.44$ bits/symbol, regardless of the impulsive channel parameters $A$ and $r$. From Fig.~\ref{fig:airG}, we can observe that as $\Lambda$ increases, the channel transitions into an IN-limited regime, where $A$ and $r$ become crucial in determining the AIR. A non-monotonic behavior of the AIR over $\Lambda$ is shown for channels with correlation parameter $r<0.9$ in Fig.~\ref{fig:airG}, where there exist inflection points where a minimum AIR occurs. Before the inflection point, the AIR quickly worsens as $\Lambda$ increases. As shown in Fig.~\ref{fig:probG0}, the IN state increases the uncertainty of the overall likelihood function due to its non-negligible long tail. However, after the inflection point, the impact of the impulsive state on the overall likelihood is reduced. In particular, when $\Lambda > 10^{3}$, the AIR for all the channels converges to a constant value. This behavior can be explained by examining the likelihood function in Fig.~\ref{fig:probG-4}. As the variance of the IN state increases, its likelihood function spreads, and the probability density around the mean decreases compared to that of the BN state. Consequently, the overall likelihood of transmitting a specific symbol $x_t$ is almost entirely dominated by the BN state. It is also shown in Fig.~\ref{fig:airG} that the AIR deteriorates as the IN becomes more pronounced with increasing $A$. Besides, the inflection point shifts toward large $\Lambda$ with a higher impulsive index $A$, which can be explained by \eqref{eq:varm}. Specifically, when the IN samples are more active (larger $A$), the power of the IN must increase (larger $\Lambda$) to maintain the same noise power. On the other hand, the channel memory helps track the noise states through the transition matrix, reducing the uncertainties of the underlying IN channel and thus improving the AIR for any given value of $A$ and $\Lambda$. 

In Fig.~\ref{fig:airG_mis}, we evaluate the AIR performance under mismatched decoding scenarios where imperfect channel knowledge exists, providing a sensitivity analysis for parameter estimation inaccuracies. In all simulations, we assume that the Markov-Middleton channel is characterized by parameters $A = 0.3$, $r = 0.9$, and $W = 4$. This setup corresponds to a highly correlated, bursty IN environment, which is commonly observed in practice~\cite{Sacuto14, chen2024}. We first examine the extreme cases by assuming $\hat{W}=1$ (black dashed line) and $\hat{\Lambda}=10^{-2}$ (orange dashed line). This simplification effectively reduces the receiver to an AWGN detector. While they perform adequately in the BN-limited regime ($\Lambda < 10^{-1}$), their performance severely deteriorates as the true channel transitions into the IN-limited regime ($\Lambda > 10^{-1}$). Additionally, ignoring channel memory by assuming $r=0$ (green dashed line) leads to a memoryless receiver design. This simplification results in an AIR loss of up to $0.1$~bits/symbol compared to the memory-aware receiver under highly correlated impulsive conditions, a finding that aligns with previous observations in~\cite{Dario09}. This highlights the need for memory-aware detection schemes or alternative techniques, such as orthogonal frequency division multiplexing (OFDM) approaches, to mitigate the impact of temporally correlated channels. Moreover, a mismatch in the impulsive parameters $A$ and $\Lambda$ imposes a notable performance loss, emphasizing the importance of accurate channel estimation in practical systems. On the other hand, the receiver is relatively robust to state mismatch: assuming a simple $2$-state model (crosses) yields similar AIR performance to the true $4$-state model. This suggests that the complexity of the receiver can be reduced without substantial performance loss as long as the key statistical properties of the IN are adequately captured.

As a reference for our practical receiver design, we present the AIR as a function of SNR in Fig.~\ref{fig:air_snr}. Specifically, the SNR required to achieve a target AIR of $1$ bits per QPSK symbol, which corresponds to a half-rate encoder, is $0.9$~dB, $2.4$~dB, and $4.2$~dB for impulsive index $A=0.1 \text{, }0.3\text{, and } 0.5$, respectively. These performance bounds will be used to benchmark our turbo-DE-PSK-IN receiver designs in the following subsection.

\begin{figure}[t]
    \centering
    \subfloat[$A=0.1$\label{fig:ber_a01}]{
        \resizebox{0.33\linewidth}{!}{\begin{tikzpicture}
	
	\begin{semilogyaxis}[
		axis line style = thick,
		grid = both,
		name = p1,
		xmin = 0.5, xmax=7,
		ymin = 7e-5, ymax=3e-1,
        xtick = {2,4,6,8},
		font=\footnotesize,
		ylabel = BER, xlabel = SNR (dB),
		legend style={
			font=\tiny,
			nodes={scale=1.0},
		},
		legend cell align={left},
		legend pos = south west,
		]
        \addplot[color=Black!40, thick, dashed] 
             table[x expr=\thisrowno{0}, y expr=\thisrowno{1}] 
            {BER_QPSK_A01N4R9L1.txt};
	
        \addplot[color=Black!60, thick, dashed] 
             table[x expr=\thisrowno{0}, y expr=\thisrowno{2}] 
            {BER_QPSK_A01N4R9L1.txt};
            
        \addplot[color=Black!80, thick, dashed] 
             table[x expr=\thisrowno{0}, y expr=\thisrowno{3}] 
            {BER_QPSK_A01N4R9L1.txt};
		
        \addplot[color=Black, thick, dashed] 
             table[x expr=\thisrowno{0}, y expr=\thisrowno{11}] 
            {BER_QPSK_A01N4R9L1.txt};

        \addplot[color=RoyalBlue!40, thick, mark=*, mark options={solid,fill=white}, mark size=2.6] 
             table[x expr=\thisrowno{0}, y expr=\thisrowno{1}] 
            {BER_DQPSK_A01N4R9L1_I1.txt};
  
        \addplot[color=RoyalBlue!60, thick, mark=*, mark options={solid,fill=white}, mark size=2.6] 
             table[x expr=\thisrowno{0}, y expr=\thisrowno{1}] 
            {BER_DQPSK_A01N4R9L1_I2.txt};

        \addplot[color=RoyalBlue!80, thick, mark=*, mark options={solid,fill=white}, mark size=2.6] 
             table[x expr=\thisrowno{0}, y expr=\thisrowno{1}] 
            {BER_DQPSK_A01N4R9L1_I3.txt};

        \addplot[color=RoyalBlue, thick, mark=*, mark options={solid,fill=white}, mark size=2.6] 
             table[x expr=\thisrowno{0}, y expr=\thisrowno{1}] 
            {BER_DQPSK_A01N4R9L1_I11.txt};

        \addplot[color=RoyalBlue!40, dotted, thick, mark=x, mark options={solid}, mark size=2.3] 
             table[x expr=\thisrowno{0}, y expr=\thisrowno{2}] 
            {BER_DQPSK_A01N4R9L1_I1.txt}; 

        \addplot[color=RoyalBlue!60, dotted, thick, mark=x, mark options={solid}, mark size=2.3] 
             table[x expr=\thisrowno{0}, y expr=\thisrowno{2}] 
            {BER_DQPSK_A01N4R9L1_I2.txt}; 

        \addplot[color=RoyalBlue!80, dotted, thick, mark=x, mark options={solid}, mark size=2.3] 
             table[x expr=\thisrowno{0}, y expr=\thisrowno{2}] 
            {BER_DQPSK_A01N4R9L1_I3.txt}; 

        \addplot[color=RoyalBlue, dotted, thick, mark=x, mark options={solid}, mark size=2.3] 
             table[x expr=\thisrowno{0}, y expr=\thisrowno{2}] 
            {BER_DQPSK_A01N4R9L1_I11.txt};

        \addplot[color=Red, thick] 
             table[x expr=\thisrowno{0}, y expr=\thisrowno{1}] 
            {BER_DQPSK_Lopt_A01N4R9L1_I31.txt}; 

        \addplot[color=Red, very thick, dashed] 
             table[x expr=\thisrowno{0}, y expr=\thisrowno{1}] 
            {BER_QPSK_LDPC_A01N4R9L1.txt}; 

        \addplot[ultra thick, color=Red, dotted] 
            coordinates {(0.9,  7e-5) (0.9, 5e-1)};
        
        \draw[<->, thick, black] (axis cs:1.95,1e-4) -- (axis cs:6.2,1e-4) node[pos=0.35, above] {$4.5$~dB};
        \draw[<->, thick, black] (axis cs:0.9,1e-4) -- (axis cs:1.95,1e-4) node[midway, above] {$1$~dB};
	\end{semilogyaxis}
 

\end{tikzpicture}    }
    }
    \subfloat[$A=0.3$\label{fig:ber_a03}]{
        \resizebox{0.33\linewidth}{!}{\begin{tikzpicture}
	
	\begin{semilogyaxis}[
		axis line style = thick,
		grid = both,
		name = p1,
		xmin = 1.8, xmax=12.3,
		ymin = 7e-5, ymax=3e-1,
		xtick = {4, 7, 10, 13},
		font=\footnotesize,
		ylabel = BER, xlabel = SNR (dB),
        legend to name=commonlegend2,
		legend style={
			font=\tiny,
			column sep=2pt,
            legend columns = 5,
			nodes={scale=1.0},
		},
		legend cell align={left},
		legend pos = south west,
		]
        \addplot[color=Black!40, thick, dashed] 
             table[x expr=\thisrowno{0}, y expr=\thisrowno{1}] 
            {BER_QPSK_A03N4R9L1.txt}; \addlegendentry{QPSK I0};
	
        \addplot[color=Black!60, thick, dashed] 
             table[x expr=\thisrowno{0}, y expr=\thisrowno{2}] 
            {BER_QPSK_A03N4R9L1.txt};\addlegendentry{QPSK I1};
            
        \addplot[color=Black!80, thick, dashed] 
             table[x expr=\thisrowno{0}, y expr=\thisrowno{3}] 
            {BER_QPSK_A03N4R9L1.txt};\addlegendentry{QPSK I2};
		
        \addplot[color=Black, thick, dashed] 
             table[x expr=\thisrowno{0}, y expr=\thisrowno{11}] 
            {BER_QPSK_A03N4R9L1.txt};\addlegendentry{QPSK I10};

        \addplot[color=Red, very thick, dashed] 
             table[x expr=\thisrowno{0}, y expr=\thisrowno{1}] 
            {BER_QPSK_LDPC_A03N4R9L1.txt}; \addlegendentry{QPSK-LDPC};
            
        \addplot[color=RoyalBlue!40, thick, mark=*, mark options={solid,fill=white}, mark size=2.6] 
             table[x expr=\thisrowno{0}, y expr=\thisrowno{1}] 
            {BER_DQPSK_A03N4R9L1_I1.txt};\addlegendentry{DE-QPSK I0};
            
        \addplot[color=RoyalBlue!60, thick, mark=*, mark options={solid,fill=white}, mark size=2.6] 
             table[x expr=\thisrowno{0}, y expr=\thisrowno{1}] 
            {BER_DQPSK_A03N4R9L1_I2.txt};\addlegendentry{DE-QPSK I1};

        \addplot[color=RoyalBlue!80, thick, mark=*, mark options={solid,fill=white}, mark size=2.6] 
             table[x expr=\thisrowno{0}, y expr=\thisrowno{1}] 
            {BER_DQPSK_A03N4R9L1_I3.txt};\addlegendentry{DE-QPSK I2};
            
        \addplot[color=RoyalBlue, thick, mark=*, mark options={solid,fill=white}, mark size=2.6] 
             table[x expr=\thisrowno{0}, y expr=\thisrowno{1}] 
            {BER_DQPSK_A03N4R9L1_I11.txt};\addlegendentry{DE-QPSK I10};

        \addplot[color=Red, thick] 
            table[x expr=\thisrowno{0}, y expr=\thisrowno{1}] 
           {BER_DQPSK_Lopt_A03N4R9L1_I31.txt}; \addlegendentry{LLopt I30};

        \addplot[color=RoyalBlue!40, dotted, thick, mark=x,mark options={solid}, mark size=2.3] 
             table[x expr=\thisrowno{0}, y expr=\thisrowno{2}] 
            {BER_DQPSK_A03N4R9L1_I1.txt};\addlegendentry{DE-QPSKsub I0};

        \addplot[color=RoyalBlue!60, dotted, thick, mark=x,mark options={solid}, mark size=2.3] 
             table[x expr=\thisrowno{0}, y expr=\thisrowno{2}] 
            {BER_DQPSK_A03N4R9L1_I2.txt};\addlegendentry{DE-QPSKsub I1};            

        \addplot[color=RoyalBlue!80, dotted, thick, mark=x,mark options={solid}, mark size=2.3] 
             table[x expr=\thisrowno{0}, y expr=\thisrowno{2}] 
            {BER_DQPSK_A03N4R9L1_I3.txt};\addlegendentry{DE-QPSKsub I2};

        \addplot[color=RoyalBlue, dotted, thick, mark=x,mark options={solid}, mark size=2.3] 
             table[x expr=\thisrowno{0}, y expr=\thisrowno{2}] 
            {BER_DQPSK_A03N4R9L1_I11.txt};\addlegendentry{DE-QPSKsub I10};

        \addplot[ultra thick, color=Red, dotted] 
            coordinates {(2.4,  7e-5) (2.4, 5e-1)};\addlegendentry{Theoretical};

	\end{semilogyaxis}
 

\end{tikzpicture}    }
    }
    \subfloat[$A=0.5$\label{fig:ber_a05}]{
        \resizebox{0.33\linewidth}{!}{\begin{tikzpicture}
	
	\begin{semilogyaxis}[
		axis line style = thick,
		grid = both,
		name = p1,
		xmin = 3.5, xmax=13.5,
		ymin = 7e-5, ymax=3e-1,
        xtick = {5, 8, 11, 14},
		font=\footnotesize,
		ylabel = BER, xlabel = SNR (dB),
		legend style={
			font=\tiny,
			nodes={scale=1.0},
		},
		legend cell align={left},
		legend pos = south west,
		]
        \addplot[color=Black!40, thick, dashed] 
             table[x expr=\thisrowno{0}, y expr=\thisrowno{1}] 
            {BER_QPSK_A05N4R9L1.txt};
	
        \addplot[color=Black!60, thick, dashed] 
             table[x expr=\thisrowno{0}, y expr=\thisrowno{2}] 
            {BER_QPSK_A05N4R9L1.txt};
            
        \addplot[color=Black!80, thick, dashed] 
             table[x expr=\thisrowno{0}, y expr=\thisrowno{3}] 
            {BER_QPSK_A05N4R9L1.txt};

        \addplot[color=Black, thick, dashed] 
             table[x expr=\thisrowno{0}, y expr=\thisrowno{11}] 
            {BER_QPSK_A05N4R9L1.txt};

        \addplot[color=RoyalBlue!40, thick, mark=*, mark options={solid,fill=white}, mark size=2.6] 
             table[x expr=\thisrowno{0}, y expr=\thisrowno{1}] 
            {BER_DQPSK_A05N4R9L1_I1.txt};

        \addplot[color=RoyalBlue!60, thick, mark=*, mark options={solid,fill=white}, mark size=2.6] 
             table[x expr=\thisrowno{0}, y expr=\thisrowno{1}] 
            {BER_DQPSK_A05N4R9L1_I2.txt}; 

        \addplot[color=RoyalBlue!80, thick, mark=*, mark options={solid,fill=white}, mark size=2.6] 
             table[x expr=\thisrowno{0}, y expr=\thisrowno{1}] 
            {BER_DQPSK_A05N4R9L1_I3.txt};

        \addplot[color=RoyalBlue, thick, mark=*, mark options={solid,fill=white}, mark size=2.6, mark repeat =2] 
             table[x expr=\thisrowno{0}, y expr=\thisrowno{1}] 
            {BER_DQPSK_A05N4R9L1_I11.txt}; 

        \addplot[color=RoyalBlue!40, dotted, thick, mark=x,mark options={solid}, mark size=2.3] 
             table[x expr=\thisrowno{0}, y expr=\thisrowno{2}] 
            {BER_DQPSK_A05N4R9L1_I1.txt};
        \addplot[color=RoyalBlue!60, dotted, thick, mark=x,mark options={solid}, mark size=2.3] 
             table[x expr=\thisrowno{0}, y expr=\thisrowno{2}] 
            {BER_DQPSK_A05N4R9L1_I2.txt};
        \addplot[color=RoyalBlue!80, dotted, thick, mark=x,mark options={solid}, mark size=2.3] 
             table[x expr=\thisrowno{0}, y expr=\thisrowno{2}] 
            {BER_DQPSK_A05N4R9L1_I3.txt};
          
        \addplot[color=RoyalBlue, dotted, thick, mark=x,mark options={solid}, mark size=2.3, mark repeat =2] 
             table[x expr=\thisrowno{0}, y expr=\thisrowno{2}] 
            {BER_DQPSK_A05N4R9L1_I11.txt};

        \addplot[color=Red, thick] 
             table[x expr=\thisrowno{0}, y expr=\thisrowno{1}] 
            {BER_DQPSK_Lopt_A05N4R9L1_I31.txt};

        \addplot[color=Red, very thick, dashed] 
             table[x expr=\thisrowno{0}, y expr=\thisrowno{1}] 
            {BER_QPSK_LDPC_A05N4R9L1.txt}; 

        \addplot[ultra thick, color=Red, dotted] 
            coordinates {(4.2,  7e-5) (4.2, 5e-1)};
	\end{semilogyaxis}
 

\end{tikzpicture}    }
    }
    \vspace{1ex}
    \pgfplotslegendfromname{commonlegend2} 

    \caption{BER performance over iterations for turbo-QPSK-IN (dashed lines), turbo-DE-QPSK-IN (circle solid lines), and suboptimal turbo-DE-QPSK-IN designs (cross dotted lines) over the Markov--Middleton channel with impulsive index (a)~$A=0.1$, (b)~$A=0.3$, and (c)~$A=0.5$, where $W=4$, $\Lambda=10$, and $r=0.9$ are fixed for all simulations. The dotted red vertical lines indicate the simulated theoretical bound derived in Fig.~\ref{fig:air_snr}. The dashed red lines show the QPSK-IN demapper with the LDPC code presented in \cite{Dario09, MMA}. The solid red lines indicate the ultimate performance bound of the proposed turbo-DE-PSK-IN receiver with perfect noise state information. All simulations have an interleaver depth of $64800$ bits.}
    \label{fig:ber}
\end{figure}

\begin{figure}[t]
    \centering
    \resizebox{!}{0.5\textwidth}{\begin{tikzpicture}
	
	\begin{semilogyaxis}[
		axis line style = thick,
		grid = both,
		name = p1,
		xmin = 1.8, xmax=17,
		ymin = 7e-5, ymax=2e-1,
		font=\footnotesize,
		ylabel = BER, xlabel = SNR (dB),
		legend style={
			font=\tiny,
			nodes={scale=1.0},
		},
		legend cell align={left},
		legend pos = south west,
		]
        \addplot[color=RoyalBlue, thick, mark=*, mark options={solid,fill=white}, mark size=2.6] 
             table[x expr=\thisrowno{0}, y expr=\thisrowno{1}] 
            {misBER_i50000.txt};\label{misber_n4l50}
	
        \addplot[color=RoyalBlue, thick, mark=x, only marks, mark size=2.3] 
             table[x expr=\thisrowno{0}, y expr=\thisrowno{2}] 
            {misBER_i50000.txt};\label{misber_n2l50}
            
        \addplot[color=RoyalBlue!70, thick, mark=*, mark options={solid,fill=white}, mark size=2.6] 
             table[x expr=\thisrowno{0}, y expr=\thisrowno{1}] 
            {misBER_i2000.txt};\label{misber_n4l2}
	
        \addplot[color=RoyalBlue!70, thick, mark=x, only marks,  mark size=2.3] 
             table[x expr=\thisrowno{0}, y expr=\thisrowno{2}] 
            {misBER_i2000.txt};\label{misber_n2l2}  

        \addplot[color=RoyalBlue!40, thick,  mark=*, mark options={solid,fill=white}, mark size=2.6] 
             table[x expr=\thisrowno{0}, y expr=\thisrowno{1}] 
            {misBER_i400.txt};\label{misber_n4l04}
	
        \addplot[color=RoyalBlue!40, thick, mark=x,  only marks, mark size=2.3] 
             table[x expr=\thisrowno{0}, y expr=\thisrowno{2}] 
            {misBER_i400.txt};\label{misber_n2l04}   
            
    \addplot[color=Black, dashed, thick, mark=*, mark options={solid,fill=white}, mark size=2.6] 
    table[x expr=\thisrowno{0}, y expr=\thisrowno{1}] 
   {misBER_PSK_i50000.txt};\label{misber_psk_n4l50}

    \addplot[color=Black, thick, mark=x, only marks, mark size=2.3] 
        table[x expr=\thisrowno{0}, y expr=\thisrowno{3}] 
    {misBER_PSK_i50000.txt};\label{misber_psk_n2l50}

    \addplot[color=Black!70, dashed, thick, mark=*, mark options={solid,fill=white}, mark size=2.6] 
        table[x expr=\thisrowno{0}, y expr=\thisrowno{1}] 
    {misBER_PSK_i2000.txt};\label{misber_psk_n4l2}

    \addplot[color=Black!70, thick, mark=x, only marks,  mark size=2.3] 
        table[x expr=\thisrowno{0}, y expr=\thisrowno{3}] 
    {misBER_PSK_i2000.txt};\label{misber_psk_n2l2}  

    \addplot[color=Black!40, dashed, thick,  mark=*, mark options={solid,fill=white}, mark size=2.6] 
        table[x expr=\thisrowno{0}, y expr=\thisrowno{1}] 
    {misBER_PSK_i400.txt};\label{misber_psk_n4l04}

    \addplot[color=Black!40, thick,mark=x,  only marks, mark size=2.3] 
        table[x expr=\thisrowno{0}, y expr=\thisrowno{3}] 
    {misBER_PSK_i400.txt};\label{misber_psk_n2l04}

	\end{semilogyaxis}
 
   \matrix[
    matrix of nodes,
    anchor=north east,
    fill = white,draw,
    inner sep = 0.2em,
    column sep = 0.1em,
    node font=\scriptsize,
      ]
      at ([xshift=-2pt, yshift=-2pt]current axis.north east){
        $W \backslash N_{s}$  & $100000$   & $4000$ & $800$\\ 
        DE-QPSK4 & \ref{misber_n4l50}  & \ref{misber_n4l2} & \ref{misber_n4l04}\\
        DE-QPSK2 & \ref{misber_n2l50}  & \ref{misber_n2l2} & \ref{misber_n2l04}\\
        QPSK4 & \ref{misber_psk_n4l50}  & \ref{misber_psk_n4l2} & \ref{misber_psk_n4l04}\\
        QPSK2 & \ref{misber_psk_n2l50}  & \ref{misber_psk_n2l2} & \ref{misber_psk_n2l04}\\
    };

\end{tikzpicture}    }
    \caption{BER performance for practical receiver design considerations with underestimated channel state $\hat{W}=2$ and limited interleaver depths of $100000$, $4000$, and $800$, respectively, for the proposed suboptimal turbo-DE-QPSK receiver after $10$ iterations. The IN channel is characterized by $W=4$, $A=0.3$, $\Lambda=10$, and $r=0.9$.}
    \label{fig:bermis}
\end{figure}
\subsection{Proposed Turbo-DE-PSK-IN Receiver Performance}
This section reports the numerically simulated bit error rate (BER) performance of the proposed optimal and suboptimal turbo-DE-PSK-IN receivers under a $4$-state Markov-Middleton IN channel. A standard half-rate convolutional code with generators $(5,7)_8$ was used. A QPSK modulation is employed to map the interleaved coded bits to symbols, after which a differential modulator processes the symbols as in \eqref{eq:diff_mod}. The differentially modulated symbols are transmitted through a Markov-Middleton channel to generate the observed sequence $\mathbf{y}_1^T$ according to \eqref{eq:y_dqpskin}. For the conventional turbo-PSK-IN receiver, the observed symbols are generated via \eqref{eq:airio}. 
For benchmarking, we include the performance of a state-of-the-art PSK-IN receiver employing the LDPC code, as demonstrated in \cite{Dario09, MMA}. To ensure a fair comparison with the DVB-S2 LDPC standard, we set the codeword length (interleaver depth) for all schemes to $64800$ bits.

Figs.~\ref{fig:ber_a01}--\ref{fig:ber_a05} show the BER performance of the proposed joint turbo-DE-PSK-IN receiver (solid lines) with respect to the number of iterations for different impulsive indices $A=0.1$, $0.3$, and $0.5$. When comparing to the performance of the conventional turbo-PSK-IN receiver (dashed lines), we observe that the turbo-DE-PSK-IN receiver significantly outperforms the turbo-PSK-IN receiver in all cases after the first turbo iteration, with a performance gain of around $4.5$~dB at a BER of $10^{-4}$ for $A=0.1$. Notably, despite the use of a relatively simple convolutional code, the proposed receiver approaches the performance of the high-complexity DVB-S2 LDPC-based PSK-IN receiver (dashed red lines) after $10$ iterations. This result highlights the efficacy of the rate-one differential structure in providing significant time diversity in impulsive environments. Furthermore, the proposed receiver maintains SNR gaps of only $1.1$ to $2.1$~dB relative to the theoretical bounds derived in Fig.~\ref{fig:air_snr}. As shown in Figs.~\ref{fig:trellis_IN2QPSK} and \ref{fig:trellis_IN2DQPSK}, these gains are achieved without increasing the trellis state complexity of the joint IN detector.


The performance of the joint turbo-DE-PSK-IN receiver is also compared to its suboptimal separate design (dotted cross lines in Figs.~\ref{fig:ber_a01}--\ref{fig:ber_a05}), where the turbo decoder and the IN detector (equalizer) are designed separately. The separate design is implemented by using the same MAP equalizer as in \cite{MMA,Alam20}, which only computes the posterior symbol probability once without iterative updates. The results show that the suboptimal separate design can closely approximate the optimal joint design without much degradation, especially in a mild IN channel (Fig.~\ref{fig:ber_a01}). Such observation was also identified in \cite{Dario09}. This finding suggests that the use of a suboptimal design is preferable as it has only around half the complexity of the optimal design for an iteration number of $10$. However, in a severely activated IN channel (Fig.~\ref{fig:ber_a05}), the separate design shows a performance loss of around $0.2$~dB compared to the joint design, which is expected to increase with a larger impulsive index $A$.

To gauge the ultimate performance of our proposed turbo-DE-PSK-IN scheme under the Markov-Middleton channel, we simulate the performance of the proposed turbo-DE-PSK-IN receiver with perfect noise state information (NSI) at the receiver. This is achieved by assuming that the noise state realization is perfectly known at the receiver, which allows for a perfect equalization of the received signal. The results are shown in Figs.~\ref{fig:ber_a01}--\ref{fig:ber_a05} as solid red lines. The perfect NSI simulation serves as a benchmark for the performance of our proposed turbo system, where we assume that a large number of iterations ($30$ iterations) are performed. The results show that the ultimate performance of the proposed turbo-DE-PSK-IN system is around $1$~dB away from the theoretical bounds, regardless of channel conditions. For $A=0.1$, a turbo-DE-PSK-IN receiver with $10$ iterations is already close to the ultimate performance, while for $A=0.5$, more iterations are required to achieve the best possible result.

In Fig.~\ref{fig:bermis}, we evaluate the performance of the separate receiver design in a setting with limited interleaver depth and mismatched channel states. The results are shown for a fixed $4$-state Markov Middleton model with $A=0.3$, $r=0.9$, and $\Lambda=10$. The number of turbo iterations is set to $10$. We first design a receiver that only assumes $\hat{W}=2$, which leads to a total of $2M=8$ trellis states in the IN detector. In this case, the receiver is designed to ignore the higher impulsive noise states. The results show that the performance of the receiver with a mismatched $\hat{W}=2$ state assumption is very close to that of the matched receiver design with $W=4$ states, which is also identified in Fig.~\ref{fig:airG_mis}. This is an important observation as we can design an efficient receiver with a reduced number of states without much performance loss. The results of Fig.~\ref{fig:bermis} also show that the receiver is sensitive to the interleaver depth. A smaller interleaver depth of $4000$ leads to a performance loss of around $2.7$~dB, while a depth of $800$ results in significant performance degradation of around $10$~dB loss at a BER of $10^{-4}$ when compared to a system with an interleaver depth of 100000 for the turbo-DE-PSK-IN system. The benefits of using a large interleaver depth are mainly twofold. First, it improves the MAP-based IN detector (equalizer) performance, and second, it effectively spreads the decoding errors for the two constituent decoders, which is crucial for the turbo decoding performance. We can see that the conventional turbo-PSK-IN (black dashed lines) can only acquire the detector-related benefits, while the proposed turbo-DE-PSK-IN system (blue solid lines) can benefit from both. This is because the turbo-DE-PSK-IN system can effectively spread the decoding errors thanks to the recursive nature of the differential decoder, which is crucial for turbo decoding performance.

\section{Conclusions}\label{sec:conc}
In this paper, we propose robust turbo receiver structures that can effectively mitigate the impact of bursty impulses, modeled by a $4$-state Markov-Middleton model, by incorporating a rate-one differential decoder into the receiver design.
We started by investigating the AIR between the PSK-modulated symbols and the channel output sequence for various channel conditions. We found that the impulsive index significantly influences the AIR. Also, we observed a non-monotonic relationship between the estimated AIR and the impulsive-to-background noise power ratio, which can be further explained by evaluating the likelihood function for the BN and IN states. Furthermore, channel memory reduces the uncertainty of the given IN channel, thereby increasing the AIR. Mismatched decoding scenarios are also considered in the AIR analysis. In the second part of the paper, we carefully derive the proposed turbo-DE-PSK-IN receiver design, which incorporates a differential decoder into the MAP-based IN detector. Results show that the proposed turbo-DE-PSK-IN receiver significantly outperforms the conventional MAP-based turbo-PSK-IN receiver by more than $4.5$~dB with an ultimate performance gap of around $1$~dB from the AIR-derived performance bound. Additionally, we proposed a suboptimal separate receiver design, which only has half the complexity of the joint design but still sufficiently approximates the optimal joint receiver design. The proposed receiver design is also shown to be robust against the underestimated IN-state scenario. 

This paper demonstrates that a simple rate-one differential encoder can effectively improve receiver performance under a serially concatenated turbo structure in the presence of bursty impulsive noise. While the present study focuses on PSK modulation and convolutional codes to clearly isolate the benefits of the proposed turbo-DE-PSK-IN architecture, the framework is inherently extensible to other modulation formats and alternative coding schemes. For instance, employing more powerful codes such as LDPC could further enhance the quality of the extrinsic information exchanged during turbo iterations. This would potentially accelerate the convergence rate, allowing the receiver to approach theoretical information limits with fewer iterations. This remains a compelling direction for future research.

\section*{Acknowledgments}
This work was funded by the RAISE collaboration framework between Eindhoven University of Technology and NXP, including a PPS-supplement from the Dutch Ministry of Economic Affairs and Climate Policy.



\end{document}